\documentclass[aps,pra,twocolumn,amsmath,amssymb,showpacs,floatfix,superscriptaddress]{revtex4-1}

\usepackage{times}
\usepackage[utf8x]{inputenc}
\usepackage[english]{babel}
\usepackage[T1]{fontenc}
\usepackage{lmodern}
\usepackage{amssymb}
\usepackage{amsmath}
\usepackage{amsthm}
\usepackage[pdftex]{graphicx}
\usepackage{epstopdf}
\usepackage{braket}
\usepackage[bottom]{footmisc}
\usepackage{dcolumn}
\usepackage{bm}
\usepackage{color}
\usepackage{relsize,dsfont,mathrsfs,empheq,verbatim,upgreek}
\usepackage{etoolbox}
\usepackage[caption = false]{subfig}
\usepackage{hyperref}
\hypersetup{
           breaklinks=true,   
        }

\usepackage{enumerate}

\newcommand{\mathd}{\mathrm{d}}
\newcommand{\Tr}{\mathrm{Tr}}
\newcommand{\Lt}{\mathcal{L}}

\begin{document}


\title{Thermodynamic Roles of Quantum Environments: From Heat Baths to Work Reservoirs}


\author{Alessandra Colla}

\affiliation{Institute of Physics, University of Freiburg, 
Hermann-Herder-Stra{\ss}e 3, D-79104 Freiburg, Germany}
\affiliation{Dipartimento di Fisica ``Aldo Pontremoli'', Universit\`a degli Studi di Milano, Via Celoria 16, I-20133 Milan, Italy}

\author{Heinz-Peter Breuer}

\affiliation{Institute of Physics, University of Freiburg, 
Hermann-Herder-Stra{\ss}e 3, D-79104 Freiburg, Germany}

\affiliation{EUCOR Centre for Quantum Science and Quantum Computing,
University of Freiburg, Hermann-Herder-Stra{\ss}e 3, D-79104 Freiburg, Germany}

\begin{abstract}
Environments in quantum thermodynamics usually take the role of heat baths. These baths are Markovian, weakly coupled to the system, and initialized in a thermal state. Whenever one of these properties is missing, standard quantum thermodynamics is no longer suitable to treat the thermodynamic properties of the system that result from the interaction with the environment. 
Using a recently proposed framework for open system quantum thermodynamics which is valid for arbitrary couplings and non-Markovian effects, we show that within the very same model, described by a Fano-Anderson Hamiltonian, the environment can take three different thermodynamic roles: a standard heat bath, exchanging only heat with the system, a work reservoir, exchanging only work, and a hybrid environment, providing both types of energy exchange. The exact role of the environment is determined by the strength and structure of the coupling, and by its initial state. The latter also dictates the long time behaviour of the open system, leading to thermal equilibrium for an initial thermal state and to a nonequilibrium steady state when there are displaced environmental modes.
\end{abstract}

\date{\today}

\maketitle

\section{Introduction}

To model the nonequilibrium dynamics of small quantum systems in contact with heat baths, and to assess their thermodynamic properties, one makes use of the theory of open quantum systems \cite{Breuer2007,Alicki1979,Kosloff2013}. This allows to link dynamical processes such as dissipation and decoherence to the exchange of heat and work between system and bath, and to the observation of irreversible behaviour in terms of positive entropy production rates \cite{Spohn1978,Lebowitz1978}.
A heat bath interacting with the system is in these cases phenomenologically modelled through a Lindblad master equation describing the evolution of the quantum system of interest, which typically leads to its thermalization to an equilibrium state described by a Gibbs state at the temperature of the bath.
The heat bath thus exchanges heat with the system while external work protocols -- represented as a time-dependent bare system Hamiltonian -- describe work performed on the system.
The employment of the Lindblad master equation to describe the dynamics stands on heavy assumptions regarding the nature of the environment and its relationship to the system: the bath must be weakly coupled to the system, it must interact so that its two-point correlation functions decay sufficiently fast with respect to the typical system relaxation time and such that the secular approximation is allowed. More so, only specific Lindblad master equations, like the weak-coupling quantum optical master equation, satisfy the detailed balance condition required for thermalization \cite{Alicki1976,Alicki2007} -- which, furthermore, requires neglecting the Lamb shift term in the system Hamiltonian.

By now, the field of quantum thermodynamics has evolved so rapidly that more and more strong coupling effects are taken into account, leading to, e.g., studies on dynamical renormalization of thermodynamic quantities \cite{Colla2022a,Huang2022} and on extensions of the famous Jarzynski equality to this regime \cite{Talkner2020}. To understand the actual role of environments for regimes where the Lindblad approach is prohibited, one needs more refined open system tools, along with a microscopic modelling of the quantum environment. 

While one expects that infinitely large environments initialized in a thermal state are proper microscopic models of a heat bath -- as can be usually argued in the Markovian regime --, it is still unclear whether strong coupling or non-Markovian effects can spoil this assumption.
For the sake of generality, one might suppose that a general environment, particularly strongly coupled and/or not memoryless, will not simply exchange heat with the system: without further assumptions, the interaction with the environment may lead to effective driving on the system, and thus to work exchange. 

In this work, we define as extreme cases a ``quantum heat bath'' as an environment which exchanges only heat with the quantum system of interest, while we term ``work reservoir'' any environment exchanging energy only in the form of work. We investigate the thermodynamic role of an infinitely large environment in the strong coupling, non-Markovian regime by examining a paradigmatic Hamiltonian -- sometimes called the Fano-Anderson model -- which is exactly solvable for the reduced system and at the same time allows a continuum of modes as the environment. The model, which is physically relevant and has several applications in condensed matter physics, nuclear, atomic and molecular physics, as well as quantum optics (see, e.g., \cite{Lambropoulos2000,Imry2002,Haug2007,Miroshnichenko2010,Lei2012}), is also suitable to represent a large heat bath from a microscopic perspective, and has all the required properties -- e.g., approach to a unique equilibrium steady state -- to describe a proper thermodynamical situation. By evaluating work and heat in the strong coupling regime according to the renormalized system Hamiltonian that emerges form the exact interaction with the environment (as suggested in Ref.\cite{Colla2022a} and, for the Fano-Anderson model, also in Ref. \cite{Huang2022}), we will see that the shape and strength of the coupling, as well as the possible initial displacement of the bath modes, play a decisive role in determining the main thermodynamic function of the environment, which in the most cases exchanges both heat and work with the system.

The structure of this work is the following. In Sec.~\ref{sec:DEQT} we recall the definitions of thermodynamic quantities according to the approach developed in \cite{Colla2022a}. These definitions are based on the canonical form of the exact time-local master equation for the system degrees of freedom.
In Sec.~\ref{sec:FA-model} we show the exact master equation for the Fano-Anderson model in terms of a general, initially uncorrelated, Gaussian state of the environment. We show that, while the model is completely relaxing to a final steady state (or to a nonequilibrium steady state (NESS) in the case of initial displacement in the environment), the environment will in general exchange both work and heat with the system even when initialized in a thermal state. In Sec.~\ref{sec:FA-extremes} we show instances where the two extremes of perfect heat bath (Sec.~\ref{sec:FA-heat}) and work reservoir (Sec.~\ref{sec:FA-work}) are reached, 
showcasing how the approach of \cite{Colla2022a} describes within the same formalism the loss of information due to the contact with a heat bath and the emergence of coherent driving of a system from a microscopic perspective. In Sec.~\ref{sec:FA-hybrid} we show and give an interpretation to a specific case where the environment acts as a hybrid agent, exchanging both work and heat with the reduced system.

In all that follows, the initial state of the system is assumed to be a completely arbitrary state. We will, however, care a great deal about the initial state of the environment; for ease of exposition, we might therefore denote the initial environmental state as simply ``the initial state'' throughout this work.

\section{Dynamically Emergent Quantum Thermodynamics}\label{sec:DEQT}

The framework for the description of the thermodynamics of an open system in the regime of general couplings, following \cite{Colla2022a}, assumes that the open system $S$ and the environment $E$, which represents a general environment, are coupled through the total microscopic Hamiltonian 
\begin{equation} \label{eq:ham-total}
 H(t) = H_S(t) + H_E + H_I(t),
\end{equation}
where $H_S$ and $H_E$ are the system and the environmental Hamiltonian, respectively, and
$H_I$ denotes the interaction between them. The time-dependency of system and interaction Hamiltonian describe possible work protocols on the system and modulation of the interaction -- including switching on and off --, while the environmental Hamiltonian is assumed to be time-independent in agreement with the assumption that we do not directly act on the environment. The total system thus undergoes a unitary evolution generated by \eqref{eq:ham-total}, which conventionally starts with an uncorrelated state between system and environment, i.e. $\rho_{SE}(0)= \rho_S(0)\otimes\rho_E(0)$.

In the strong coupling regime, the interaction energy associated to the Hamiltonian $H_I$ is non-negligible with respect to the bare system energy associated to $H_S$. This energy surplus results in a renormalization of the system energy levels -- as it happens, for example, with the Lamb-shift induced by the electromagnetic vacuum in the weak-coupling regime. In the general case, as developed in \cite{Colla2022a}, the renormalized Hamiltonian for the system emerges from the dynamics and is found as the unitary part of the generator for the open system, i.e. the operator $K_S$ in the canonical form of the exact time-local master equation for the open system $S$ (also known as time-convolutionless master equation) \cite{Hall2014,Breuer2012}:
\begin{equation} \label{eq:tcl-meq}
 \frac{d}{dt}\rho_S(t) = \Lt_t[\rho_S(t)] = -i \left[K_S(t),\rho_S(t)\right] + {\mathcal{D}}_t  [\rho_S(t)].
\end{equation}
The second part of the master equation, also called dissipator, is of the following form
\begin{equation} \label{eq:dissipator}
 {\mathcal{D}}_t [\rho_S] = \sum_{k}\gamma_{k}(t)\Big[L_{k}(t)
  \rho_S L_{k}^{\dag}(t) - \frac{1}{2}\big\{L_{k}^{\dag}(t)L_{k}(t),\rho_S\big\}\Big]
\end{equation}
with decay rates $\gamma_k(t)\in \mathbb{R}$ and corresponding Lindblad operators $L_k(t)$. The above general structure of the time-local master equation can be
derived from the requirements of the preservation of Hermiticity and trace of the density matrix 
\cite{Hall2014,Breuer2012}. 
This differential equation, which, being exact, is given in general by an explicitly time-dependent 
generator $\Lt_t$, completely describes the dynamics of the system and thus how all strong coupling and memory effects influence it \cite{Shibata1977,Shibata1979}. While it in principle exists for any open system \cite{Colla2022a}, its exact derivation, namely its explicit expression in terms of Hamiltonian parameters and initial environmental state,
is often challenging. Algorithmic methods when the exact derivation is not feasible include perturbation expansions in the system-environment coupling through the time-convolutionless projection operator technique \cite{Shibata1977,Shibata1979,Breuer2007,vanKampen1974a,vanKampen1974b}, and numerical techniques such as HEOM \cite{Gatto2024}.
In the present paper, the model under study is exactly solvable and the master equation can be derived exactly, see \cite{Tu2008,Jin2010,Lei2012,Zhang2012} and Sec.~\ref{sec:FA-model}.

The separation of the master equation \eqref{eq:tcl-meq} into a dissipator and a part generating unitary evolution is not unique \cite{Breuer2007,Hall2014}, but can be made so by imposing tracelessness of the Lindblad operators $L_k(t)$ \cite{Hall2014,Sorce2022,Colla2022a}. This requirement is equivalent to minimizing the dissipator part as a superoperator according to a set of norms \cite{Sorce2022}, including one that is averaging over input and output pure random states following the Haar measure on the unitary group, giving equal weight to all pure states of the open system. Alternatively, it means that all the non-Hermitian Lindblad operators are taken to describe coupling between orthogonal states, so that their dissipator terms identify processes linked to heat exchange.

Via the identification of the emergent Hamiltonian $K_S(t)$, which can also in principle be time-dependent, as a renormalized energy operator for the system, the internal energy of the system is defined as
\begin{equation} \label{eq:internal-energy}
 U_S (t) = \Tr \{K_S(t)\rho_S(t)\}.
\end{equation}
The first law of thermodynamics is then given by
\begin{eqnarray}\label{eq:first-law}
 \Delta U_S (t) \equiv U_S(t) - U_S(0) = W_S(t) + Q_S(t),
\end{eqnarray}
where the work and heat contributions arise from the change in energy levels and in the system state, respectively:
\begin{eqnarray}
 W_S(t) &=& \int_0^t d\tau \, \Tr \big\{ \dot{K}_S(\tau) \rho_S(\tau) \big\},  \label{work} \\
 Q_S(t) &=& \int_0^t d\tau \, \Tr \big\{ K_S(\tau) \dot{\rho}_S(\tau)  \big\}.  \label{heat}
\end{eqnarray}
Because of the structure of \eqref{eq:tcl-meq}, the heat contribution $Q_S(t)$ is determined by the change in the system state due to the dissipative evolution, as it can be expressed as the following
\begin{equation}
 Q_S(t) = \int_0^t d\tau \, \Tr \big\{ K_S(\tau) \mathcal{D}_{\tau}[\rho_S(\tau)] \big\}.
\end{equation}

Assuming furthermore that the bath is sufficiently large such that it does not suffer noticeable temperature changes during the evolution, entropy production is defined in analogy with the Clausius inequality as
\begin{eqnarray}\label{eq:DEQT-ep-bare}
 \Sigma_S (t) = \Delta S_S(t) - \beta \delta Q_S(t) \;,
\end{eqnarray}
with $\Delta S_S(t)=S(\rho_S(t))-S(\rho_S(0))$ the change of the von Neumann entropy of the reduced system and $\beta$ the inverse initial temperature of the environment.
Under the condition that the instantaneous Gibbs state of the reduced system $\rho_S^G(t) = e^{-\beta K_S(t)}/Z_S(t)$ suffers no dissipation, i.e.  $\mathcal{D}_t [\rho_S^G(t)]=0 $, one can prove \cite{Colla2022a} that negative entropy production rates imply the presence of non-Markovian effects in the sense of information backflow \cite{Breuer2009,Breuer2016,Wissmann2015}
\begin{eqnarray}
\sigma_S(t) \equiv \dot{\Sigma}_S(t) \leq 0  \implies  \text{non-Markovianity} ,
\end{eqnarray}
and one may therefore understand information backflow as a necessary condition for breaking the second law of quantum thermodynamics.

\section{The Fano-Anderson model}\label{sec:FA-model}

We consider the integrable model of a bosonic mode, our reduced system, linearly coupled to a continuum of bosonic modes, the environment, within rotating wave approximation. This model, which can be equivalently formulated in terms of fermionic modes, is known as the Fano-Anderson model \cite{Fano1961,Anderson1961,Mahan2000} and describes quantum systems coupled to environment through particle exchange. The Hamiltonian of the total system reads
\begin{equation}\label{eq:FA-total-Hamiltonian}
H^{\text{FA}} = \underbrace{\omega_0 a^\dag a}_{H_S} +  \underbrace{\sum_j \omega_j c_j^\dag c_j}_{H_E} +  \underbrace{\sum_j (g_j a^\dag c_j + g_j^* a c_j^\dag)}_{H_I} \; , 
\end{equation}
where $a^\dag$, $a$ are the bosonic creation and annihilation operators on the central system, whose main frequency is $\omega_0$, and $c_j^\dag$, $c_j$ are the creation and annihilation operators of the environmental modes with frequencies $\omega_j$. The parameters $g_j$ represent the coupling constants between the central mode $a$ and each bath mode $c_j$.

\subsection{Exact time-local master equation}\label{sec:FA-TCL}
For a Gaussian initial state of the environment, the reduced dynamics is Gaussianity preserving and can be solved exactly by solving the Heisenberg equations of motion \cite{Zhang2012,Mahan2000}. Then, the exact TCL master equation can be found by making an ansatz for its generator, and comparing the coefficients with the exact evolution of first and second moments of the system \cite{Xiao2013,Picatoste2024}.
Assuming that the modes in the environment are initially prepared uncorrelated and each in a Gaussian state, the exact master equation for the central mode in the Fano-Anderson model reads (see Appendix~\ref{app:FA-master-equation} for the derivation):
\begin{align}\nonumber
\frac{d}{dt}\rho_S(t) =& -i \left[\omega_r(t) a^{\dagger}a +i f(t) a^\dag -i f^*(t)a,\rho_S(t)\right] \\
 &  + \gamma(t)(N(t)+1) \left[a\rho_Sa^{\dagger}-\frac{1}{2}\left\{a^{\dagger}a,\rho_S\right\}\right] 
 \nonumber \\
& + \gamma(t) N(t) \left[a^{\dagger}\rho_Sa-\frac{1}{2}\left\{aa^{\dagger},\rho_S\right\}\right] \nonumber \\ 
    &  + \delta^*(t)\left[a^\dag\rho_Sa^{\dagger}-\frac{1}{2}\left\{a^{\dagger}a^\dag,\rho_S\right\}\right] \nonumber\\
  &  + \delta(t) \left[a\rho_Sa-\frac{1}{2}\left\{aa,\rho_S\right\}\right] 
 \label{eq:FA-TCL}
  \;.
\end{align}
All the coefficients of the master equation are dependent on the coupling strength and structure, and in general on the initial temperature, displacement and squeezing of the environment. In the above, the parameters $\omega_r(t)$ and $\gamma(t)$  depend exclusively on the coupling through the Green function $G(t)$ solving 
\begin{equation}\label{eq:FA-green-function}
\dot{G}(t) +i \omega_0 G(t) +\int_0^t d\tau \mathcal{K}(t-\tau)G(\tau) =0 \; ,
\end{equation}
with $G(0)=1$, where $ \mathcal{K}(t-\tau) = \sum_j |g_j|^2 e^{-i\omega_j (t-\tau)}$ is the memory kernel due to the coupling landscape, i.e. depending on the spectral density $J(\omega)$ in the continuum limit 
\begin{equation} \label{eq:FA-memory-kernel-2}
 \mathcal{K}(t -\tau) = \int_0^{\infty} d\omega J(\omega) e^{-i\omega (t-\tau)} \; .
\end{equation}
The two parameters are then given explicitly by
\begin{align}
\omega_r(t) = - \Im\left\{\frac{\dot{G}(t)}{G(t)}\right\} \; , \quad
\gamma(t) = - 2\Re\left\{\frac{\dot{G}(t)}{G(t)}\right\} \; .
\end{align}

All other parameters in the master equation depend on the initial state of the environment. For example, the parameter $\delta(t)$ is due to initial squeezing of the environment and reads
\begin{equation}
\delta(t) = \dot{J}(t) - 2\frac{\dot{G}(t)}{G(t)}J(t)  \; ,
\end{equation}
with
\begin{equation}
J(t) = \sum_j g_j^2 \langle\langle c_j c_j \rangle\rangle_0 \left[ \int_0^t d\tau G(t-\tau)e^{-i\omega_j\tau}\right]^2 \; ,
\end{equation}
where we have defined the notation
\begin{equation}
\langle\langle AB \rangle\rangle_0 := \langle AB \rangle_0 - \langle A\rangle_0\langle B\rangle_0 \; .
\end{equation}
The other coefficients are also affected by initial squeezing; however, for zero squeezing we have that $\delta(t)$ vanishes (and with it the last two terms in \eqref{eq:FA-TCL}), and that the other parameters take on specific meanings.

The coefficient $N(t)$, for example, depends then only on the initial temperature, describes the available number of excitations in the environment as seen from the system (see Sec.~\ref{sec:FA-thermo}), and reads
\begin{equation}
N(t) = I(t) + \frac{\dot{I}(t)}{\gamma(t)} \; ,
\end{equation}
with noise integral
\begin{equation}
I(t) = \sum_j |g_j|^2 \langle\langle c_j^\dag c_j \rangle\rangle_0 \left| \int_0^t d\tau G(t-\tau)e^{-i\omega_j\tau}\right|^2 \; .
\end{equation}
It vanishes for the zero temperature case.

The parameter $f(t)$ represents instead the emergence of coherent driving and depends only on the initial displacement of the modes
\begin{equation}
f(t) = \dot{F}(t) - \frac{\dot{G}(t)}{G(t)}F(t) 
\end{equation}
with
\begin{equation}
F(t) := -i \sum_j g_j \langle c_j \rangle_0 \int_0^t d\tau G(t-\tau) e^{-i\omega_j\tau} \; .
\end{equation}
The driving force $f(t)$ thus vanishes in the case of no initial displacement of the environmental modes. We remark that no external driving of this form was imposed on the bare system Hamiltonian $H_S$, and that the linear driving in the master equation is entirely emerging from tracing out the environmental degrees of freedom. It is useful to note that, were we to impose such an external linear driving on the bare system, it would in principle undergo renormalization due to coupling with the (thermal) environment, unless the latter is in some sense inert, see Appendix~\ref{app:FA-driving}.

The master equation is already in generalized Lindblad form and satisfies the principle of minimal dissipation, as the Lindblad operators $a$ and $a^\dag$ can be considered traceless even if the Hilbert space is infinite (they are traceless in any truncation of the Hilbert space to any number of Fock states). In the case of a thermal state, it had been already obtained in \cite{Tu2008,Jin2010,Lei2012} using path integral techniques, and in \cite{Picatoste2024} using the present technique. 

\subsection{Effective Hamiltonian, heat and work}\label{sec:FA-thermo}
According to Sec.~\ref{sec:DEQT} we can thus identify the emergent Hamiltonian for the Fano-Anderson model from the coherent part of the master equation, which gives
\begin{equation} \label{eq:FA-K_S}
 K_S(t) = \omega_r(t) a^{\dagger}a -i f(t) a^\dag +i f^*(t)a \;.
\end{equation}
It represents a bosonic mode with a renormalized, time-dependent frequency, and a time-dependent non-adiabatic driving term with force $f(t)$ (which is not periodic in the general case). The driving term arises exclusively in the presence of initial displacement in the bath, while the time-dependent renormalization of the frequency is a general feature, appearing also at zero temperature. 
Notice that since $\omega_r(t)$ depends only on the Green function $G(t)$, then it depends exclusively on the bare central frequency $\omega_0$ and on the spectral density $J(\omega)$. We remark that this is not a typical case in open quantum systems as $K_S(t)$ depends in general also on, e.g, the environmental temperature (see the case of the Jaynes-Cummings model \cite{Smirne2010,Colla2022a}). We ascribe this effect as due to the Gaussianity of the evolution.

For the rest of this work, we focus on the case where the bath is in general thermal, and possibly displaced, such that $\delta(t)=0$. This allows us to explore different roles of the environment as a heat bath or a work reservoir. Before we do that, though, we argue that the Hamiltonian \eqref{eq:ham-total} is in principle suitable as a microscopic model of a mode coupled to a thermal bath; we do so by looking at the general expression for first law quantities -- namely work, heat and internal energy -- in the case of a thermal initial state of the environment, and showing that the open system relaxes to an equilibrium state in the long time limit (in the thermal case) or to a nonequilibrium steady state (NESS) in the case of additional initial displacement of the environment.

Considering the initial environmental state to be thermal, $\rho_E(0)= e^{-\beta H_E}/Z_E$, leads to vanishing parameters in the master equation \eqref{eq:FA-TCL}, namely $f(t)=\delta(t) \equiv 0$. Then, the effective Hamiltonian is $K_S(t)= \omega_r(t)a^\dag a$ and the internal energy of the central mode follows simply as
\begin{equation} \label{eq:FA-internal-energy}
 U_S(t) = \omega_r(t) n(t) \;,
\end{equation}
where we denoted with $n(t):= \langle a^\dag a \rangle_t$ the evolved average excitation number of the system. Consequently, the net work exchanged up to time $t$ is given by
\begin{equation} \label{eq:FA-work}
 \delta W_S(t) = \int_0^t d \tau \dot{\omega}_r(\tau) n(\tau) \;,
\end{equation}
with a work exchange rate $\dot{\omega}_r(t) n(t)$ that is present only for a renormalized frequency $\omega_r$ that varies in time. The net heat exchanged is instead given by
\begin{equation} \label{eq:FA-heat-basic}
 \delta Q_S(t) = \int_0^t d \tau {\omega}_r(\tau) \dot{n}(\tau) \;.
\end{equation}
Using the master equation \eqref{eq:FA-TCL} we find that $ \dot{n}(t) = \gamma(t) [N(t) - n(t)]$, and we can divide heat into two distinct contributions
\begin{align} \label{eq:FA-heat}\nonumber
 \delta Q_S(t) &= \int_0^t d \tau [ \dot{Q}_S^{\text{in}}(\tau) - \dot{Q}_S^{\text{out}}(\tau)  ] \\
 &= \int_0^t d \tau {\omega}_r(\tau) \gamma(\tau) [N(\tau) - n(\tau)] \;.
\end{align}
Here,  $\dot{Q}_S^{\text{out}}(t) =  {\omega}_r(t) \gamma(t) n(t)$ describes the outgoing heat rate, and is given by the energy gain (gained, in this case, by the environment) $\omega_r(t)$ times the renormalized system-bath transition rate $\gamma(t)$ and the average excitations $n(t)$ in the system; this part of the heat therefore describes how outgoing heat exchange arises from excitations in the central mode spontaneously decaying to the environment. In a similar way, $\dot{Q}_S^{\text{in}}(t) =  {\omega}_r(t) \gamma(t) N(t)$ describes the incoming heat rate (this time gained by the system) where we find the same renormalized frequency $\omega_r(t)$ and the same transition rate $\gamma(t)$. What changes is the available number of excitations, given by $N(t)$,  which as mentioned previously assumes the role of a renormalized average excitation number of the bath, representing the available excitations present in the environment as seen from the system. Note that this does not mean that $N(t)$ is the actual average number of excitations in the bath (this will in general depend on the initial state of the system), but it represents in a way the number of excitations that are ``available'' to enter into the system from the environment. In the Born-Markov limit \cite{Breuer2007}, for example, $N(t)$ becomes time-independent and approaches the Planck distribution at the initial temperature of the environment.

\subsection{Relaxation to equilibrium}\label{sec:FA-relaxation}

Let us show that, in the continuum limit, the open system in the Fano-Anderson model (with a thermal initial environmental state) reaches a unique equilibrium state.
To witness complete relaxation of the reduced system -- namely that all initial states converge to a unique steady state at long times -- we simply assume that the Green function $G(t)$ decays to zero at infinity
\begin{equation}\label{eq:FA-G-decay}
G(t) \xrightarrow[t\rightarrow \infty]{} 0 \; .
\end{equation}
This leads to the disappearance of any initial condition contribution to all the relevant moments, see eqs. \eqref{eq:FA-moment-a}-\eqref{eq:FA-moment-a*a}. In particular we find vanishing expectation values for $a$ and $aa$, and a final excitation number given by the noise integral:
\begin{align} 
 &\langle a \rangle_t  \xrightarrow[t\rightarrow \infty]{} 0 ,\label{eq:FA-moment-a-long} \\
  &\langle aa \rangle_t  \xrightarrow[t\rightarrow \infty]{} 0 , \label{eq:FA-moment-aa-long} \\
 &\langle a^{\dagger}a \rangle_t  \xrightarrow[t\rightarrow \infty]{}   I(\infty) =: \overline{n}. \label{eq:FA-moment-a*a-long}
\end{align}
The above implies that the unique final steady state is thermal,
\begin{equation}
\overline{\rho}_S = \frac{e^{-X a^\dag a}}{\Tr\{e^{-X a^\dag a}\}} \; , \quad \text{with} \; \; X= \ln{\left( \frac{1+\overline{n}}{\overline{n}}\right)} \; .
\end{equation}
The entire noise integral $I(\infty)$, namely the final excitation number, can be written explicitly in terms of the spectral density $J(\omega)$ (see Appendix~\ref{app:FA-relaxation} for the proof):
\begin{equation}\label{eq:FA-noise-infty-3-main}
\overline{n} = \int_0^\infty \mathd \omega J(\omega) \frac{1}{e^{\beta \omega}-1}   \frac{1}{[\omega_0+\Delta(\omega)-\omega]^2+\pi^2 J^2(\omega)}  \; ,
\end{equation}
where we defined the principal value integral
\begin{equation}
\Delta(\omega) = \mathcal{P} \int_0^\infty \mathd \omega' \frac{J(\omega')}{\omega - \omega'} \; .
\end{equation}
The final steady state of the central mode in the case of a thermal environment is thus determined by both the initial temperature and the spectral density.

Using this result, it is possible to prove that the final steady state also coincides with the expected mean-force equilibrium state, namely the partial trace over the environment of a global thermal state at the initial temperature of the bath:
\begin{equation}
\rho_S^*= \Tr_E \left\{ \frac{e^{-\beta H_{SE}}}{Z_{SE}}\right\}  \;.
\end{equation}
It is important to note that such a statement should always be proven and can never be assumed: since the total system is a closed system undergoing unitary evolution, and since it was initialized in a product state, it is impossible for it to reach a global state $e^{-\beta H_{SE}}/{Z_{SE}}$. The fact that it looks this way from the open system perspective as a result of information loss must be shown for the specific model. The full proof makes use of the fluctuation dissipation theorem -- and the assumption that all poles of $\hat{G}(z)$ (with $\hat{G}$ the Laplace transform of $G(t)$) have negative real part -- and is reported in Appendix~\ref{app:FA-relaxation}.

Under these assumptions, and that in the long-time limit also the parameters $\omega_r$ and $\gamma$ reach a unique final value, the effective Hamiltonian converges to $\overline{K}_S= \overline{\omega}_r a^\dag a$. A natural question is how the steady state emergent Hamiltonian relates to the Hamiltonian of mean force \cite{Campisi2009HMF}, namely the temperature-dependent Hamiltonian $H^*_S$ such that
\begin{equation}
\frac{e^{-\beta H^*_{S}(\beta)}}{Z^*_{S}}= \Tr_E \left\{ \frac{e^{-\beta H_{SE}}}{Z_{SE}}\right\} \; .
\end{equation}
While they are indeed related, it turns out that they are not the same; the Hamiltonian of mean force depends on temperature, while $K_S$, in this model, does not. They are instead connected by the relation
\begin{equation}\label{eq:equiv-MFH}
\beta H^*_{S} = \overline{\beta}_r \overline{K}_S + \text{const.} \; ,
\end{equation}
where $\overline{\beta}_r$ is the steady state value of a renormalized temperature $\beta_r(t)$, which connects the instantaneous equilibrium state of the master equation \eqref{eq:FA-TCL} with a Gibbs state relative to $K_S(t)$. The instantaneous renormalized temperature can also help in formulating a second law of quantum thermodynamics for the Fano-Anderson model, whose violations have a clear-cut connection to the presence of information backflow. We will illustrate this next.

\subsection{Instantaneous Gibbs state at the renormalized temperature}\label{sec:FA-renorm-temp}

We define the time-dependent renormalized temperature $\beta_r(t)$ directly from the parameters in the master equation \eqref{eq:FA-TCL} via
\begin{equation}
\gamma (t) N(t) = \gamma (t) (N(t)+1) e^{\beta_r(t) \omega_r(t)} \; ,
\end{equation}
which leads to 
\begin{equation}
\beta_r(t) = \frac{1}{\omega_r(t)}\ln\left(\frac{N(t)+1}{N(t)} \right) \; .
\end{equation}
Notice that this also means that the coefficient $N(t)$, which we interpreted in Sec.~\ref{sec:FA-thermo} as the number of excitations available in the environment for the system to absorb, is given in terms of the renormalized temperature by
\begin{equation}
N(t) = \frac{1}{e^{\beta_r(t) \omega_r(t)} -1} \; ,
\end{equation}
namely by a Planck distribution for the renormalized frequency $\omega_r(t)$ and temperature $\beta_r(t)$. This leads to the interpretation of $\beta_r(t)$ as an effective temperature of the environment, as perceived by the system through the interaction. 

Using this renormalized temperature, we define the following instantaneous Gibbs state 
\begin{equation}
\rho_S^{G_r}(t) = \frac{e^{-\beta_r(t) \omega_r(t) a^\dag a}}{\Tr\{e^{-\beta_r(t) \omega_r(t) a^\dag a} \}} \; .
\end{equation}
It is easy to prove that, due to the structure of the master equation in the thermal case, this is a fixed point of the evolution, in parallel with Markovian master equations satisfying detailed balance \cite{Alicki1976,Alicki1987}. Namely, $\Lt_t [\rho_S^{G_r}(t)] = 0 $ for all times, with $\Lt_t$ the time local generator describing the master equation \eqref{eq:FA-TCL}. 

We exploit this property to formulate a fully renormalized second law of quantum thermodynamics for the Fano-Anderson model, by defining the entropy production for the system as
\begin{equation}
\Sigma_S(t) = \Delta S(\rho_S(t)) - \int_0^t d\tau \beta_r(\tau) \dot{Q}_S(\tau) \; ,
\end{equation}
i.e. we have modified Eq. \eqref{eq:DEQT-ep-bare} in order to take into account a varying temperature of the environment, as it is seen from the perspective of the open system through the interaction. 
We do this in the same way as in \cite{Strasberg2021}, though with a different definition of bath temperature.
This leads to a form of entropy production rate
\begin{equation}
\sigma_S(t) = -\frac{d}{dt}\Bigg|_{\tau=0}S(\rho_S(t+\tau)||\rho_S^{G_r}(t))\; ,
\end{equation}
where, however, the new instantaneous Gibbs state is automatically, at each point in time, a fixed point of the evolution. This form of entropy production rate, thus, is such that violations of the second law of thermodynamics (namely, $\sigma_S(t)< 0$ for some $t$), imply the presence of information backflow, as shown in \cite{Colla2022a}. The difference here is that, using the renormalized temperature, no additional condition is needed to establish the link between information flow and entropy production rates.

Furthermore, regarding approach to equilibrium in the long time limit, we find automatically that the system steady state coincides with the renormalized Gibbs state 
\begin{equation}
\overline{\rho}_S^{G_r} = \overline{\rho}_S \; .
\end{equation}
From this it follows that the renormalized temperature in the long time limit is also such that 
\begin{equation}
\braket{a^\dag a}_{\infty} = \frac{1}{e^{\overline{\beta}_r \overline{\omega}_r} -1} \; ,
\end{equation}
showing that the average occupation number of the system oscillator approaches in the long time limit a Planck distribution with renormalized frequency and temperature. Thus, $\overline{\beta}_r$ also takes the role as the effective temperature of the system, indicating how system and environment have reached equilibrium. The renormalized temperature in the relaxation limit indeed coincides with the system temperature (at infinite times) that was defined for the Fano-Anderson model in \cite{Huang2022}.

Moreover, recalling the fact that the final steady state is the mean force state $\rho_S^*$, it is clear that the long time limit of the emergent Hamiltonian $K_S$ in this model is linked to the Hamiltonian of mean force via relation \eqref{eq:equiv-MFH}.

\subsection{Strong coupling nonequilibrium steady state}\label{sec:FA-NESS}

In the case where there is some initial displacement of the environmental modes, the system does not reach thermal equilibrium, as it is continuously driven out of it by the driving term in the Hamiltonian. However, under the same assumption \eqref{eq:FA-G-decay} of vanishing Green function in the long time limit, the system asymptotically approaches a unique NESS described by the long time limit of the moments
\begin{align} 
 &\langle a \rangle_t  \sim \overline{F}(t) \;,\label{eq:FA-moment-a-long-NESS} \\
  &\langle aa \rangle_t  \sim  \overline{F}^2(t) \; , \label{eq:FA-moment-aa-long-NESS} \\
 &\langle a^{\dagger}a \rangle_t \sim    \overline{n} + |\overline{F}(t)|^2 
 \; , \label{eq:FA-moment-a*a-long-NESS}
\end{align}
i.e. the state evolves as the thermal state $\overline{\rho}_S^{G_r}$ from the last section, but with a time-dependent displacement (see Appendix~\ref{app:FA-NESS} for details)
\begin{equation}
\overline{F}(t) = -i \sum_j \varphi_j e^{-i\omega_j t} 
\end{equation}
with
\begin{equation}
\varphi_j=  \alpha_j g_j \hat{G}(-i\omega_j)  \; ,
\end{equation}
where $\alpha_j$ is the initial displacement of the $j$-th mode and $\hat{G}$ is the Laplace transform of $G$. We can therefore denote the nonequilibrium steady state of the open system with a displaced equilibrium state
\begin{equation}\label{eq:FA-NESS-displaced-Gibbs}
\overline{\rho}_S(t) = D_t \overline{\rho}_S^{G_r} D_t^{\dag} \; ,
\end{equation}
with 
\begin{equation}
D_t:= D[\overline{F}(t)] = e^{\overline{F}(t)a^\dag - \overline{F}^*(t)a}
\end{equation}
the displacement operator for the time-dependent displacement $\overline{F}(t)$. The master equation generator in the long time limit indeed reads 
\begin{equation}\label{eq:FA-NESS-long-time-me}
\overline{\mathcal{L}}_t[X] = -i [\overline{K}_S(t), X] + \overline{\mathcal{D}}[X] \; ,
\end{equation}
where $\overline{\mathcal{D}}$ is the dissipator with rates given by $\overline{\gamma}\overline{N}$ and $\overline{\gamma}(\overline{N}+1)$ and the (still time-dependent) effective Hamiltonian reads
\begin{equation}
 \overline{K}_S(t) = \overline{\omega}_r a^{\dagger}a +i \overline{f}(t) a^\dag -i \overline{f}^*(t)a \;,
\end{equation}
with driving force
\begin{equation}\label{eq:FA-NESS-force}
\overline{f}(t) = \sum_j \left[\overline{\omega}_r - \omega_j -\frac{i}{2}\overline{\gamma}\right]\varphi_j e^{-i\omega_j t} \; .
\end{equation}
Then, the long-time dynamics of the state $\overline{\rho}_S(t)$ becomes a unitary evolution given by
\begin{equation}\label{eq:FA-NESS-evolution}
\frac{d}{dt}\overline{\rho}_S(t) = \overline{\mathcal{L}}_t[\overline{\rho}_S(t)] \equiv \dot{D}_t D_t^{\dag}\overline{\rho}_S(t) + \overline{\rho}_S(t)D_t\dot{D}_t^{\dag} \; .
\end{equation}

We call the state $\overline{\rho}_S(t)$ a NESS because the state of the open system tends asymptotically to this unique, unitarily evolved state after the function $G(t)$ has decayed, independently of its initial conditions. We remark, however, that it does not represent a NESS in the usual sense \cite{Freitas2022,Chen2024}, as, in general, the state is not periodic in time and its energy fluxes are not constant. The latter can be seen from the fact that its internal energy
\begin{equation}
 \overline{U}_S(t) = \overline{\omega}_r \left(\overline{n} + |\overline{F}(t)|^2\right) +i (\overline{f}(t) \overline{F}^*(t) + \text{h.c.})
\end{equation}
is a non-trivial function of time. The internal energy of the state in the long time evolution thus keeps changing, along with heat and work fluxes. 

A particular case where the asymptotic state is also a NESS in the usual sense is when there is only one displaced mode in the environment. Let us denote the mode with the index $d$. Then, the displacement of the steady state is periodic in time (with period $T=2\pi/\omega_d$) and given by
\begin{equation}
\overline{F}(t) = -i \varphi_d e^{-i\omega_d t} = -i\alpha_d g_d \hat{G}(-i\omega_d) e^{-i\omega_d t} \; .
\end{equation}
The generator $\overline{\mathcal{L}}_t$ of the master equation then assumes the form of the Floquet-Lindblad master equation \cite{Blumel1991,Breuer1997} which describes the dynamics of periodically driven dissipative quantum systems.
We note at this point that the result \eqref{eq:FA-NESS-displaced-Gibbs} has been obtained earlier in \cite{Louisell1965} by different methods. 

In the case of a single-mode driving, the internal energy of the system approaches the constant value
\begin{equation}
 \overline{U}_S(t) = \overline{\omega}_r \left(\overline{n} - |\varphi_d|^2\right) + 2\omega_d |\varphi_d|^2 \; ,
\end{equation}
which is in general a function of renormalized temperature $\overline{\beta}_r$, of the frequency, displacement and coupling of the driving mode ($\omega_d$, $\alpha_d$, $g_d$), but also of the entire spectral density $J(\omega)$ describing the coupling to the rest of the environment.

Since there is no variation of internal energy in the long-time limit, the fluxes of work and heat (which are themselves constant) balance each other out:
\begin{align}
 \dot{\overline{Q}}_S =& - \dot{\overline{W}}_S \nonumber\\ 
  =& -\overline{\gamma}\omega_d|\varphi_d|^2 \; ,
\end{align}
so that there is work consistently being performed on the system to keep it out of equilibrium at a constant energy. Moreover, since the long time evolution is unitary, the von Neumann entropy of the system also reaches the constant value of $S(\overline{\rho}_S^{G_r})$. The entropy production rate is thus determined solely by the heat rate and gives a measure of how far the NESS is from thermal equilibrium:
\begin{align}\label{eq:FA-NESS-ep}
\dot{\sigma} =& -\overline{\beta}_r\dot{\overline{Q}}_S(t)\nonumber \\ 
 =& \overline{\beta}_r \overline{\gamma} \omega_d|\alpha_d|^2|g_d|^2|\hat{G}(-i\omega_d)|^2 \; .
\end{align}
Here the Laplace transform of the function $G$ reads, in the continuum limit (see Appendix \ref{app:FA-relaxation-eq})
\begin{equation}
\hat{G}(-i\omega_d) = \frac{1}{i(\omega_0+\Delta(\omega_d)-\omega_d)+\pi J(\omega_d)} \; ,
\end{equation}
where we defined the principal value integral
\begin{equation}
\Delta(\omega_d) = \mathcal{P} \int_0^\infty \mathd \omega' \frac{J(\omega')}{\omega_d - \omega'} \; .
\end{equation}
The denominator of $\hat{G}(-i\omega_d)$ shows a generalized resonance condition for the system and the driving mode. 
Therefore, the entropy production for the NESS becomes stronger when the displaced mode in the environment and the system are closer to resonance. Furthermore, typically, the higher the driving mode coupling and displacement, the farther the NESS from equilibrium.

\section{Extreme roles of the environment: heat bath vs. work reservoir}\label{sec:FA-extremes}

Now that we have seen the general solution, master equation and thermodynamic features of the Fano-Anderson model \eqref{eq:FA-total-Hamiltonian}, we want to discuss, in this section, the impact of details like coupling strength, spectral density shape, and initial environmental state on the thermodynamic properties of the central mode. In particular, since we have seen in Sec.~\ref{sec:FA-model} that arbitrary coupling to the external environment leads in general to both heat and work exchange, we 
want to define special cases that induce extreme thermodynamic regimes -- namely, situations in which the environment acts as either a proper heat bath, exchanging only heat with the central mode, or as a work reservoir, changing only the energy levels of the system in time with negligible dissipation, thus exchanging only work-like energy.

We start in Sec.~\ref{sec:FA-heat} by considering the Born-Markov limit of the master equation, in the weak coupling regime, and showing that there is no extra work contribution given by the environment, while it is in general responsible for non-zero heat exchange.  We see that this is also the case of a purely Markovian bath modelled through a completely flat spectral density, without the need for perturbation theory, showcasing how purely white-noise, Markovian environments take the role of heat baths. On the contrary, we present in Sec.~\ref{sec:FA-work} a situation where the displaced initial state of the environment induces driving -- thus, an emergent work protocol -- on the system while the contribution coming from dissipation, i.e. heat exchange, is negligible with respect to the driving in the semiclassical limit. This is a case where we see clearly that tracing out a special environment -- even an infinite one endowed with a temperature -- can still give rise to an effective unitary evolution of the system which is governed by a time-dependent Hamiltonian.

\subsection{Perfect heat bath: Born-Markov limit and white noise}\label{sec:FA-heat}

We enter the Born-Markov regime first by assuming weak system-environment coupling. We thus isolate a coupling (perturbation) parameter $\lambda$ from the interaction Hamiltonian, which then becomes $\lambda H_I(t)$. We then determine the second order expansion of the parameters $\gamma(t)$ and $\omega_r(t)$ (see Appendix~\ref{app:FA-Born-Markov}). This last quantity, in particular, is still time dependent, showcasing that a second order expansion in the coupling still leads to time dependency in the emergent Hamiltonian, and thus to work exchange in general. As a consequence, weak coupling -- in the sense of small coupling parameters -- is not enough to justify the assumption that the environment acts as a thermal bath. However, if we now also assume to be in the regime where the Markovian approximation is valid, then one obtains the following time-independent values for the renormalized frequency and rate:
\begin{align}
\gamma^{M} =& 2 \lambda^2 \pi J(\omega_0)  \; ,  \\
\omega_r^{M} =& \omega_0 + \lambda^2 \delta \omega =  \omega_0 + \lambda^2 \mathcal{P} \int_0^\infty \mathd \omega \frac{J(\omega)} {(\omega_0-\omega)}    \; .
\end{align}

The shift $\lambda^2 \delta \omega$ is the usual Lamb shift of the Lindblad master equation. Especially due to the time-independence of the renormalized frequency $\omega_r^M$, there is no extra work contribution from the bath, therefore zero work on the reduced system $\delta W_S(t) =0$, and the environment acts as a true heat bath. In the current case, the heat rate inherits an additional term due to the shift of the renormalized frequency, so that $\dot{Q}_S(t) = \Tr\{(\omega_0 + \lambda^2 \delta \omega) a^\dag a \mathcal{D}[\rho_S(t)]\}$; nonetheless, since the dissipator is already in second order in $\lambda$, this contribution is neglected and we obtain results that are compatible with the weak coupling formulation of quantum thermodynamics \cite{Alicki1995,Kosloff2013}. Moreover, this limit is recovered also if an external linear driving is imposed on the system (see Appendix~\ref{app:FA-driving}), as the approximations are such that the work protocol does not undergo renormalization due to the coupling to the bath (we remark, however, that this is a special case).

Another simple case in which the environment does not perform any work on the system, but for which 
we can avoid a perturbative treatment, is that of a completely flat spectral density describing a fully Markovian bath, namely $J(\omega) = \gamma_0/2\pi$. Note that the rate $\gamma_0$ need not be small with respect to other scales in the system, if we allow the extension of the integral in the memory kernel \eqref{eq:FA-memory-kernel-2} to the whole real axis. Then this gives
\begin{equation}
\mathcal{K}(t-\tau) =  \gamma_0 \delta(t-\tau) \; ,
\end{equation}
such that the differential equation \eqref{eq:FA-green-function} becomes, under the convention $\int_0^\epsilon \delta(x) dx = 1/2$, the following
\begin{equation}
\dot{G}(t) +i \omega_0 G(t) + \frac{\gamma_0}{2} G(t) = 0 \; ,
\end{equation}
with simple solution $G(t) = e^{-i\omega_0 t} e^{-\gamma_0 t/2}$. Then, the renormalized frequency coincides with the bare value, $\omega_r(t)\equiv \omega_0$. This means that there is no time-dependent renormalization of the system Hamiltonian, and thus no work contribution in the non-driven case of the Fano-Anderson model. Moreover, the transition rate is equivalent to the rate in the spectral density, i.e. $\gamma(t) \equiv \gamma_0$. Since this rate is time independent, it also holds that a linear driving protocol also does not get modified by the presence of the bath (see Appendix~\ref{app:FA-driving}). From the arguments above, it is also likely that there is no kind of driving protocol on the system which would undergo renormalization in this case of a purely Markovian bath, and we therefore consider this idealized case as the most paradigmatic case of a ``proper'' heat bath.

\subsection{Perfect work reservoir: displaced initial state}\label{sec:FA-work}
We now study a particular case where the environment assumes a completely different function for the system, namely that of a work reservoir performing a protocol on the central mode. 
Assume that, instead of a purely thermal state, the environment is initialized in a displaced thermal state. 
It is well known from quantum optics that a displaced electromagnetic field vacuum (zero temperature) is equivalent to a classical driving term in the system Hamiltonian \cite{Cohen1998}. Here we explore this property to construct a work reservoir at any finite temperature.

Let the displacement be arbitrary for now, and let $\vec{\alpha}= \{ \alpha_j\}_j$ denote the set of displacements that each mode in the environment $c_j$ undergoes, so that the collective displacement operator is given by 
\begin{equation}
D(\vec{\alpha}) = \bigotimes_j D_j(\alpha_j) \; , \quad \text{with} \;\; D_j(\alpha_j) = e^{ \alpha_j c_j^\dag - \alpha_j^* c_j } \; .
\end{equation}
Furthermore, we denote by $|\alpha_{\text{max}}|$ the maximum displacement magnitude, which we assume can in principle be performed on different modes of the environment. Let us label the modes with maximum displacement with the index $k$, such that $\alpha_k = |\alpha_{\text{max}}|e^{i\theta_k}$.
As we did in the previous section, we also assume weak coupling between the system and the environment by isolating the small coupling parameter $\lambda$ from the interaction Hamiltonian, such that each coupling strength between the $j$-th mode and the central oscillator is given by $\lambda g_j$.

The limit of the Fano-Anderson model as a system coupled to a work reservoir is now obtained if we take the semiclassical limit, namely vanishing coupling $\lambda \rightarrow 0$ and infinite displacement $|\alpha_{\text{max}}| \rightarrow \infty$ such that the parameter $\epsilon := \lambda |\alpha_{\text{max}}|$ is finite. 
Since the Green function $G(t)$ depends only on the coupling strength and not on the initial state of the environment, all the parameters in the master equation depending only on $G(t)$ are easily found in the weak coupling limit as follows:
\begin{align}
&G(t)  \xrightarrow[\lambda\rightarrow 0]{} e^{-i\omega_0t} , \\
&\omega_r(t)  \xrightarrow[\lambda\rightarrow 0]{} \omega_0 , \\
&\gamma(t)  \xrightarrow[\lambda\rightarrow 0]{} 0 ,
\end{align}
while the noise integral $I(t)$ depends on the initial temperature of the environment but not on the initial displacement, such that
\begin{equation}
I(t)  \xrightarrow[\lambda\rightarrow 0]{} 0 . \\
\end{equation}
This already shows that, in the limit of vanishing coupling, the dissipator in the master equation \eqref{eq:FA-TCL} vanishes completely, as expected. What survives, however, is the linear driving term. 
Indeed, rewriting the term $F(t)$ in terms of $\epsilon$, namely
\begin{equation}
F(t) = -i \epsilon \sum_j g_j \frac{\alpha_j}{|\alpha_{\text{max}}|} \int_0^t d\tau G(t-\tau) e^{-i\omega_j \tau} \; ,
\end{equation}
we can take the limit of weak coupling and large displacement, so that only the modes we previously labelled with $k$ survive, giving
\begin{equation}
F_{\text{cl}}(t) = - \epsilon \sum_k g_k e^{i\theta_k} \frac{ e^{-i\omega_k t} - e^{-i\omega_0t}}{\omega_0-\omega_k}\; .
\end{equation}
It leads to a residual driving in the emergent Hamiltonian described by
\begin{equation}
f_{\text{cl}}(t) = - i \epsilon \sum_k g_k e^{i\theta_k} e^{-i\omega_k t} \; .
\end{equation}

The dynamics of the open system in the semiclassical limit is therefore described by a master equation for a closed, driven bosonic mode, namely 
\begin{equation}\label{eq:FA-semiclassical}
\dot{\rho}^{\text{cl}}_S(t) = - i \left[ \omega_0 a^\dag a +  f_{\text{cl}}(t) a^\dag - f_{\text{cl}}^*(t) a,\rho^{\text{cl}}_S(t)\right] \; .
\end{equation}
The infinitely large environment of the Fano-Anderson model is then responsible for a work protocol on the system with negligible dissipation, and can thus be identified as an ideal work reservoir. 
If there is only one maximally displaced mode, we recover the equation for a harmonic oscillator driven by coherent light. 
We remark here that a similar conclusion for the Caldeira-Legget model -- i.e., that an initially displaced environment leads to an effective work contribution -- has been reported in \cite{Cavina2024} while this manuscript was in preparation.
Additionally, a link between the emergence of effective work and the presence of coherences in the initial state of the environment is identified in \cite{Rodrigues2019}, although this connection is established through a different framework.

\section{Hybrid environment}\label{sec:FA-hybrid}
In the previous sections, we have looked at two examples for which the environment in the Fano-Anderson model has extreme thermodynamic functions, either exchanging only work or only heat with the central mode. These examples required strong assumptions -- like that of fully Markovian behaviour or of large environmental mode displacement -- and also assumed a small coupling parameter. Deviating from these particular assumption results in the environment acting in a hybrid regime, namely exchanging \emph{both} work and heat with the central mode. Especially at stronger couplings, this represents the most general case. Nonetheless, it is beneficial to look explicitly at a specific example of a hybrid reservoir, and to try to interpret the mechanisms behind this behaviour, especially when the initial state is thermal (such that the environment would traditionally be seen as a heat bath).

We start with a thermal environmental state but choose the coupling with the environment in such a way that non-Markovian behaviour is present. This is done, in our case study, by picking a spectral density with a well defined peak; the structure of choice is that of a Lorentzian, as it makes most of the calculation completely analytic as reported also in \cite{Picatoste2024}, where the fact that the environment performs work on the system is exploited to obtain enhanced efficiencies for the Otto cycle.
We use this here instead to give an interpretation of the emergent work by understanding the main peak of the spectral density as an effective driving mode through a reaction coordinate mapping of the model. This hints at the conclusion that structured couplings -- by means of peaked spectral densities -- are responsible for non-Markovian behaviour which, in turn, leads to driving (time-dependent) contributions in the effective Hamiltonian, even when starting in a thermal state of the environment.

\subsection{Lorentzian spectral density}\label{sec:FA-Lorentzian-solution}
Let us recall from \cite{Picatoste2024} the more explicit results for the parameters in the master equation by assuming the spectral density to be a Lorentzian, i.e.
\begin{equation}\label{eq:lorentzian_spectral_density_wc}
    J(\omega) = \frac{\gamma_0}{2 \pi} \frac{\eta^2}{(\omega_c - \omega)^2 + \eta^2} \; .
\end{equation}
Here, $\eta$ represents the spectral width -- as a guiding intuitive principle, therefore, the smaller $\eta$, the more non-Markovian effects will be present -- and $\omega_c=\omega_0-\Delta$ describes the position of the peak, so that $\Delta$ gives the detuning between the central mode and the spectral density peak. The parameter $\gamma_0$, instead, gives a measure for the strength of the coupling and corresponds to the Markovian decay rate when at resonance ($\Delta=0$):
\begin{equation}
\gamma^{M} = 2\pi J(\omega_0) = \gamma_0 \frac{\eta^2}{\Delta^2+\eta^2} \; , \quad \;  \; \gamma^{M} |_{\Delta=0} = \gamma_0 \; .
\end{equation}

With the spectral density \eqref{eq:lorentzian_spectral_density_wc}, the memory kernel \eqref{eq:FA-memory-kernel-2} can be found analytically by extending the range of the integral to the entire real axis and assuming that we can neglect the effects from the fact that $J(\omega)\neq 0$ at negative frequencies (for a discussion on this point see \cite{Picatoste2024}). It is given by
\begin{equation}\label{eq:lorentzian-memory-kernel}
\mathcal{K}(t) = \frac{\gamma_0 \eta}{2} e^{-\eta |t|} e^{-i \omega_c t} \; ,
\end{equation}
such that the Green's function $G(t)$ can be found via Laplace transform of its integro-differential equation \eqref{eq:FA-green-function} as
\begin{align}\label{eq:FA-G-lorentzian}
    G(t) = \frac{e^{-i \omega_0 t}}{\mu_2 - \mu_1} \left(\mu_2 e^{\mu_1 t} - \mu_1 e^{\mu_2 t}\right) \;,
\end{align}
where $\mu_{1,2}$ are the roots of the quadratic equation 
\begin{equation}
 \mu^2 + (\eta - i \Delta) \mu + \frac{\gamma_0 \eta}{2} = 0 \;.
\end{equation}
Notice that, in this case, the Green's function \eqref{eq:FA-G-lorentzian} has the wanted properties mentioned in Sec.~\ref{sec:FA-relaxation}, which are sufficient to prove that the system always relaxes to a unique steady state, and that this state is exactly the reduced thermal equilibrium state at the initial temperature of the environment.

Then, the coupling to the bath induces a time-dependent renormalization of the oscillator frequency, which in this case reads
\begin{equation}\label{eq:FA-wr-lorentzian}
 \omega_r(t)  = \omega_0 - \text{Im} \left\{\mu_1 \mu_2 \frac{e^{\mu_1 t} - e^{\mu_2 t}}{\mu_2 e^{\mu_1 t} - \mu_1 e^{\mu_2 t}} \right\} \; ,
\end{equation}
and leads to work contributions to the reduced system's energy due to the interaction with the environment. The time-dependent transition rate $\gamma(t)$ is instead given by
\begin{equation}\label{eq:FA-gamma-lorentzian}
 \gamma(t)  =  \text{Re} \left\{\mu_1 \mu_2 \frac{e^{\mu_1 t} - e^{\mu_2 t}}{\mu_2 e^{\mu_1 t} - \mu_1 e^{\mu_2 t}} \right\} \; .
\end{equation}

From these considerations, and from the fact that both a well-defined time-dependent frequency renormalization and a non-zero transition rate exist, we conclude as expected that the environment exchanges both heat and work with the reduced system. 

\subsubsection{Second order parameters}
Evaluating main parameters of the Fano-Anderson master equation, like $\omega_r$ or $\gamma$, at second order of approximation is useful to understand how spectral density parameters like width and height influence the renormalization and dynamical properties of the central mode. In the case of a Lorentzian spectral density, we find (see Appendix \ref{app:FA-Born-Markov} for the second order expansion)
\begin{equation}
\frac{\dot{G}^{(2)}(t)}{G^{(2)}(t)} = - i \omega_0 -\frac{\gamma_0 \eta}{2} \frac{e^{(i\Delta-\eta)t}-1}{i\Delta -\eta} \; ,
\end{equation}
which leads to the following renormalized frequency and transition rate (see also \cite{Breuer2007}):
\begin{align}
   \omega_r^{(2)} (t) =& \omega_0 + \frac{\gamma_0 \eta \Delta / 2}{\eta^2 + \Delta^2} \left[1 - e^{- \eta t} \Big( \cos \Delta t + \frac{\eta}{\Delta} \sin \Delta t \Big) \right] \; ,\\
   \gamma^{(2)} (t) =&  \frac{\gamma_0 \eta^2}{\eta^2 + \Delta^2} \left[1 - e^{- \eta t} \Big( \cos \Delta t - \frac{\Delta}{\eta} \sin \Delta t \Big) \right] \; .
\end{align}
The terms feature oscillations of frequency given by the detuning $\Delta$. This is true for the second order expansion, while higher orders, as well as the exact result,  might feature instead more complex frequency spectra. From the above we conclude that, at second order, more peaked spectral densities (smaller $\eta$) have longer lived oscillations in the parameters, while wider spectral densities (larger $\eta$) converge faster to the long time limit values. For the renormalized frequency, the long time (Markovian) limit is given by
\begin{equation}\label{eq:FA-wr-lorentzian-2nd-order}
  \lim_{t\rightarrow \infty} \omega_r^{(2)} (t) =  \omega_0 + \frac{\gamma_0 \eta \Delta / 2}{\eta^2 + \Delta^2} \;.
\end{equation}
At small detuning with respect to spectral width ($\Delta/\eta \ll 1$), the renormalized frequency is higher for more peaked spectral densities, while the opposite is true for larger detuning with respect to the width ($\Delta/\eta \gg 1$); thus, the magnitude of the renormalized frequency -- at least at low orders in the coupling strength -- seems to be dictated by the slope of the spectral density at the value of the bare central frequency, as was also observed for thermodynamic quantities in \cite{Picatoste2024}. Analogously, the renormalized transition rate at infinite times, i.e.
\begin{equation}
   \lim_{t\rightarrow \infty}  \gamma^{(2)} (t) =   \frac{\gamma_0 \eta^2}{\eta^2 + \Delta^2} \; ,
\end{equation}
is instead determined by the height of the spectral density at the bare central frequency, and gives simply $\gamma_0$ at resonance, or in the limit of large $\eta$ (Markovian limit).

\subsection{Spectral density peaks do work}\label{sec:FA-Lorentzian-RC}

The intuition that the peak of the Lorentzian spectral density is responsible for the emergent work protocol, coming from the previous observation that flat spectral densities lead to a perfect heat bath (Sec.\ref{sec:FA-heat}) and from the study of second order parameters, can be better understood and interpreted by performing a reaction coordinate mapping \cite{Garg1985,Cao1997,Garraway1997,Hartmann2000,Thoss2001,Hughes2009,Martinazzo2011,Roden2012, Woods2014,Iles-Smith2014,Iles-Smith2016,Strasberg2016}. This method consists in a unitary normal mode transformation of the environment, which allows to incorporate the most important environmental degrees of freedom into a single resulting collective mode, called the reaction coordinate (RC).
The RC couples directly to the reduced system -- here, what the reduced system actually \emph{is} is not very important; the method works in several different scenarios -- and, at the same time, interacts with the residual environment. The transformed residual environment couples only to the RC and not to the original reduced system, and the hope is that this coupling is weak and such that it is possible to apply second-order Born-Markov approximations and to obtain a standard Lindblad master equation for the enlarged reduced system -- namely, the original reduced system plus the RC.

This method is efficient and widely used -- particularly also in quantum thermodynamics, see e.g. \cite{Strasberg2016} and \cite{Nazir2018} --, and is general to the point that it can also be applied to a fermionic bath \cite{Strasberg2018}. In our context we apply it to the Fano-Anderson bosonic bath, leading to a Hamiltonian for the enlarged reduced system that describes the interaction between the central mode and the RC \cite{Nazir2018}, namely
\begin{equation}\label{eq:FA-RC-Hamiltonian-S}
H_{S+RC} = \omega_0 a^\dag a +  \omega_{RC} b^\dag b + g a^\dag b + g^* a b^\dag \; ,
\end{equation}
with $b$ denoting the bosonic creation and annihilation operators for the reaction coordinate and $|g|$ the coupling strength between RC and central mode, while the RC is also weakly coupled to the residual Markovian reservoir, now identified by a bath of bosonic modes $\tilde{c}_j$:
\begin{equation}\label{eq:FA-RC-Hamiltonian}
H_{RC} = H_{S+RC} + \sum_j \omega_j \tilde{c}_j^\dag \tilde{c}_j + \sum_j\left[ \tilde{g}_j b^\dag \tilde{c}_j  + \tilde{g}_j^* b \tilde{c}_j^\dag \right]  \; ,
\end{equation}
with a modified coupling strength $|\tilde{g}_j|$. In the continuum limit, this corresponds to a different spectral density $\tilde{J}(\omega)$ with respect to the original one. 
Usually, if the original spectral density is peaked enough (such that there is one distinct main frequency in the bath that can be extracted with the RC mapping), then the residual spectral density is expected to allow a standard Born-Markov treatment of the residual bath. We will see next that this is indeed the case for a Lorentzian spectral density, as long as the bandwidth $\eta$ is not too large.

\subsubsection{Reaction coordinate mapping}
The method behind the reaction coordinate mapping is well known and we will not report it here. The mapping depends on the original coupling (which must be linear) and on the desired one as outcome of the procedure; it is possible to obtain the final Hamiltonian \eqref{eq:FA-RC-Hamiltonian} from the Fano-Anderson Hamiltonian, and the relevant mapping -- namely, the derivation of formulas for $|g|$, $\omega_{RC}$ and $\tilde{J}(\omega)$ -- can be found in \cite{Nazir2018}. 

Our choice of a Lorentzian spectral density can lead to technical problems, e.g. diverging integrals, due to its inherent unphysical nature as a spectral density. However, this choice often leads to straightforward analytical results if we extend the frequency integrals to the whole real axis, and assume that the contributions due to this are negligible. We apply the same prescription to the reaction coordinate mapping, so that values converge and are consistent. Then, the coupling parameter between system and RC is given by
\begin{equation}
|g|^2 = \int_{-\infty}^{+\infty} J(\omega) d\omega = \mathcal{K}(0) = \frac{\gamma_0 \eta}{2} \; ,
\end{equation}
and can be made arbitrarily large by increasing $\gamma_0$. The width $\eta$ also influences the coupling, but we will assume next that it represents a small parameter.
The reaction coordinate frequency $\omega_{RC}$ is given by
\begin{equation}
\omega_{RC} = \frac{1}{|g|^2}  \int_{-\infty}^{+\infty} d\omega  \omega J(\omega) = \frac{\eta}{\pi} \int_{-\infty}^{+\infty} d\omega \frac{\omega+\omega_c}{\omega^2 + \eta^2}  = \omega_c \; ,
\end{equation}
where we have evaluated the integral symmetrically between values $-\Omega$ and $\Omega$ and then taken the limit $\Omega \rightarrow + \infty$.  Thus, the RC frequency is in this case given by the frequency at which the spectral density peaks.

The new spectral density describing the coupling between the RC and the rest of the modes is then given by
\begin{equation}
\tilde{J}(\omega) =  \frac{|g|^2 J(\omega)}{\left[ \mathcal{P} \int_{-\infty}^{+\infty} d\omega' \frac{J(\omega')}{\omega' - \omega}\right]^2 + \left[ \pi J(\omega)\right]^2} = \frac{|g|^2 J(\omega)}{|\hat{\mathcal{K}}(-i \omega)|^2} \; ,
\end{equation}
where we used eq. \eqref{eq:FA-Laplace-K-iw}. With expression \eqref{eq:lorentzian-memory-kernel} for the memory kernel of a Lorentzian spectral density, we find that $|\hat{\mathcal{K}}(-i \omega)|^2 = \pi \gamma_0 J(\omega)/2$, so that the new spectral density is flat and simply given by
\begin{equation}
\tilde{J}(\omega) = \frac{\tilde{\gamma}_0}{2\pi} \; , \quad \quad \text{with} \quad \tilde{\gamma}_0 = 2\eta \; .
\end{equation}
This corresponds to a flat spectral density with coupling strength $\tilde{\gamma}_0$ proportional to the width of the original Lorentzian spectral density. This means that the more peaked the spectral density is, the lower is the resulting coupling between the RC and the residual bath. From this we conclude that for small width $\eta$ we are in the regime where the Born-Markov approximation is allowed, and we can derive a Lindblad master equation for the system+RC which fits very well the exact dynamics and reads \cite{Breuer2007}
\begin{equation}\label{eq:FA-RC-me}
\frac{d}{dt}\rho_{S+RC}(t) = -i [H_{S+RC}, \rho_{S+RC}] +\mathcal{D}[\rho_{S+RC}] \; ,
\end{equation}
where the dissipator is given by two terms with $b$ and $b^\dag$ operators for the RC as Lindblad operators, i.e.
\begin{equation}
\mathcal{D}[\rho] = W_- \left[b \rho b^\dag -\frac{1}{2} \{ b^\dag b , \rho\} \right] + W_+ \left[b^\dag \rho b -\frac{1}{2} \{ b b^\dag , \rho\} \right] \; ,
\end{equation}
and where the corresponding rates are such that 
\begin{equation}
W_- - W_+ = \tilde{\gamma}_0 = 2\eta \; .
\end{equation}

The described procedure tells us that the exact evolution of the central mode can either be obtained by solving for the system using the entire non-Markovian bath, or by tracing out the RC in the reaction coordinate mapping using the master equation \eqref{eq:FA-RC-me}.

What we see therefore is that a Fano-Anderson model with a sufficiently peaked Lorentzian spectral density is equivalent to a reaction-coordinate model where the central mode is linearly coupled to a second mode, which is in turn connected to a Markovian bath. Since the master equation for system+RC is in Lindblad form with a time-independent Hamiltonian, we know that the residual bath is not performing any work and only providing heat to the two modes. The actual work protocol on the central mode must therefore emerge from tracing out the RC. As the RC mode comes out of a well defined narrow peak in the spectral density, we obtain from this the intuition that peaks in environmental spectral densities are responsible for the emergence of driving (work) on the system.

We remark, nonetheless, that this does not mean that spectral density peaks lead only to work. Indeed, coupling a system with a single mode leads to both heat and work contributions; by taking the limit $\eta\rightarrow 0$ with $|g|^2= \gamma_0 \eta/2$ fixed, we obtain the limiting case of two coupled bosonic modes, as
\begin{equation}
\lim_{\eta\rightarrow 0} J(\omega) = |g|^2 \delta(\omega - \omega_c) \; ,
\end{equation}
while the residual bath disappears. 
Then, even assuming weak coupling such that we can look at second order quantities, we find
\begin{align}
\lim_{\eta\rightarrow 0} \omega_r^{(2)} (t) &= \omega_0 + \frac{\gamma_0 \eta}{ 2\Delta}  \left[1 - \cos \Delta t \right]  \; , \\
\lim_{\eta\rightarrow 0} \gamma^{(2)} (t) &=  \frac{\gamma_0 \eta}{\Delta}\sin \Delta t  \; , 
\end{align}
which implies both heat and work contributions.

\section{Conclusions}
In this paper, we examined the diverse roles that the same quantum environment can take by analyzing the Fano-Anderson model, for different structures of spectral density and different initial environmental states. The model, while exactly solvable, is entirely suitable for thermodynamic studies since, as we showed, it leads to relaxation to equilibrium (when the environment is initially in a thermal state) or to the approach to a NESS (if there is initial displacement in the environmental modes).
By leveraging a recently proposed framework for open system quantum thermodynamics, we were able to make considerations that go beyond standard weak-coupling quantum thermodynamics.
Crucial aspects we found for the model are the renormalization of the system frequency and the driving amplitude induced by the interaction with the environment, and the thermodynamic properties of the nonequilibrium steady state emerging in the case of periodic driving, which can be described by a
master equation in Floquet-Lindblad form.
Furthermore, we demonstrated that a single model could reveal an environment acting as a heat bath, work reservoir, or a hybrid of both. 

We found that in the regime of weak-coupling and Markovian dynamics, the environment starting from a thermal state typically exchanges only heat with the system, and thus takes the role of a heat bath as it is traditionally assumed. Outside of weak coupling, a flat spectral density is still required to obtain a similar result, showcasing how Markovian noise is related to heat exchange. On the contrary, when the environment is initialized in a displaced thermal state, it can exert coherent driving forces on the system. In the semiclassical limit of weak coupling and large displacement, the heat exchange becomes negligible and the environment exchanges only work with the system, behaving as a work reservoir. Finally, we explored the case of a hybrid environment starting from a thermal state, where yet both heat and work exchanges occur simultaneously. Choosing a Lorentzian spectral density and understanding the dynamics through a reaction-coordinate mapping, we attribute the emergence of work protocols to peaks in the spectral density. 

From this study we conclude that the actual definition of a quantum heat bath needs to be stated with caution: interactions at a quantum level, even to an infinite number of degrees of freedom, can be highly sensitive to details (like the initial state of the environment) and can turn the same physical environment from a heat bath to an effective work reservoir. In particular, we identify two sources of effective work on quantum systems: initial coherences in the environment (displaced environmental modes), and colored noise, i.e. structured spectral densities leading to non-Markovian effects.

Learning how to properly engineer a quantum environment (and accordingly, its coupling to the system) is therefore crucial for the employment of quantum thermodynamics outside of weak coupling in practical applications. Emergent work protocols can, for example, be exploited for the purpose of autonomous quantum heat engines \cite{Tonner2005,Rasola2024}, or to enhance standard thermodynamic cycles \cite{Picatoste2024}. 


\acknowledgments
We would like to thank Andrea Smirne and Franklin Rodrigues for useful feedback on the project and the manuscript. This project has received funding from the European Union's Framework Programme 
for Research and Innovation Horizon 2020 (2014-2020) under the 
Marie Sk\l{}odowska-Curie Grant Agreement No.~847471. We acknowledge support by the Open Access Publication Fund of the University of Freiburg. This work has also been supported by the German Research Foundation (DFG) through FOR 5099.


\appendix
\section{Exact TCL master equation}\label{app:FA-master-equation}
We derive here the exact master equation \eqref{eq:FA-TCL} for the Fano-Anderson model in an initial Gaussian state.
From the commutation relations
$[c_j,c^{\dagger}_k]=\delta_{jk}$, $[a,a^{\dagger}]=1$, and $[a,c^{\dagger}_j]=0$, we write the exact coupled Heisenberg equations of motion for all system and bath modes according to the total Hamiltonian \eqref{eq:FA-total-Hamiltonian}:
\begin{align}
\dot{a}(t) &= -i \omega_0 a(t) -i \sum_j g_j c_j(t) \; , \\
\dot{c}_j(t) &= -i \omega_j c_j(t) -i  g_j^* a(t) \; ,
\end{align}
which, formally solving for $c_j(t)$, lead to
\begin{equation}
c_j(t) = e^{i\omega_j}c_j(0) - i g_j^* \int_0^t \mathd \tau e^{-i\omega_j(t-\tau)} a(\tau) \; ,
\end{equation}
and thus to the integro-differential equation for the system mode
\begin{equation}\label{eq:FA-integrodiff-a-app}
\frac{d}{d t}a(t)  +i \omega_0 a(t) + \int_0^t \mathd \tau \mathcal{K}(t-\tau) a(\tau) = c(t) \; .
\end{equation}
The dynamics of the reduced system is thus determined by the coupling with the environmental modes and their initial condition as encoded in the above defined memory kernel and inhomogeneity, which read
\begin{align} \label{eq:FA-memory-kernel-app}
 \mathcal{K}(t,\tau) &\equiv  \mathcal{K}(t-\tau) 
 := \sum_j |g_j|^2 e^{-i\omega_j (t-\tau)} \; , \\ \label{eq:FA-def-c}
c(t) &:= -i \sum_j g_j e^{-i\omega_j t} c_j(0) \; .
\end{align}
We take the continuum limit in the environment by introducing a spectral density $J(\omega)$ whose role is to describe both the density of bath modes present at a certain frequency, and how strongly these modes couple to the central system. Therefore we identify
\begin{equation}\label{eq:FA-spectral-density-app}
J(\omega) = \sum_j |g_j|^2 \delta(\omega-\omega_j) \;,
\end{equation}
which can be replaced, as in all practical applications, by a continuous function. The choice of this function therefore determines the properties of the environment and its coupling with the system. In terms of the spectral density the memory kernel is then written as
\begin{equation} \label{eq:FA-memory-kernel-2-app}
 \mathcal{K}(t,\tau) = \int_0^{\infty} d\omega J(\omega) e^{-i\omega (t-\tau)} \; .
\end{equation}

To solve the integro-differential equation \eqref{eq:FA-integrodiff-a-app}, we define the Green function $G(t)$ as the solution of the homogeneous integro-differential equation
\begin{equation} \label{eq:FA-green-function-app}
 \frac{d}{dt}G(t) + i\omega_0 G(t) +  \int_0^t \mathd \tau \mathcal{K}(t-\tau) G(\tau) = 0
\end{equation}
corresponding to the initial value $G(0)=1$. From this it follows the exact Heisenberg evolution for $a$, namely
\begin{equation} \label{eq:FA-Heisenberg-solution-app}
 a(t) = G(t) a(0) + \int_0^t \mathd \tau G(t-\tau) c(\tau) \; ,
\end{equation}
which then gives all relevant information on the dynamics of the reduced system, as a function of initial conditions $a(0)$ and $c_j(0)$ only.  We assume a factorizing total initial state $\rho_{SE}(0) = \rho_S(0) \otimes \rho_E(0)$,  where the initial environmental state is for now a general Gaussian state of decoupled modes
\begin{equation} \label{eq:FA-rho-env-dec}
 \rho_E(0) = \bigotimes_{j}\rho_j(0) \; .
\end{equation}
so that
\begin{align} \label{eq:FA-moments-c}
 \langle c(t) \rangle &= -i \sum_j g_j e^{-i\omega_j t} \langle c_j(0) \rangle \;, \\
 \langle a(0) c(t) \rangle &= -i \sum_j g_j e^{-i\omega_j t} \langle a(0)\rangle\langle c_j(0)\rangle \; , \\
 \langle a(0) c^{\dagger}(t) \rangle &= -i \sum_j g_j e^{i\omega_j t} \langle a(0)\rangle\langle c_j^\dag(0)\rangle \;, \\ 
  \langle c(t_1) c(t_2) \rangle &= - \sum_{j k} g_j g_k e^{-i\omega_j t_1} e^{-i\omega_k t_2} \langle c_j(0) c_k(0)\rangle \; , \\ 
   \langle c^\dag(t_1) c(t_2) \rangle &=  \sum_{j k} g^*_j g_k e^{i\omega_j t_1} e^{-i\omega_k t_2} \langle c^\dag_j(0) c_k(0)\rangle \; .
\end{align}

From the Heisenberg evolution \eqref{eq:FA-Heisenberg-solution-app} we can find expressions for the first and second moments of the open system mode, which represent all relevant moments of the system,  in terms of the Green function $G(t)$ and the initial environmental conditions:
\begin{align} 
 \langle a \rangle_t = & G(t) \langle a \rangle_0 + F(t) \;, \label{eq:FA-moment-a} \\ \nonumber
 \langle aa \rangle_t = & G^2(t) \langle aa \rangle_0 + 2 G(t) F(t) \braket{a}_0 \\ &+F^2(t) - J(t), \label{eq:FA-moment-aa} \\ \nonumber
 \langle a^{\dagger}a \rangle_t = & |G(t)|^2 \langle a^{\dagger}a \rangle_0 + G(t)F^*(t)\braket{a}_0 \\ &+ G^*(t)F(t)\braket{a^\dag}_0 + |F(t)|^2 + I(t), \label{eq:FA-moment-a*a}
\end{align}
where we have defined, using the notation $\braket{\braket{AB}}_0 = \braket{AB}_0-\braket{A}_0\braket{B}_0$,
\begin{align} \label{eq:F}
F(t):=& -i \sum_j g_j \int_0^t d\tau  G(t-\tau)  e^{-i\omega_j \tau} \braket{c_j}_0 \; ,\\ \label{eq:FA-noise-integral}\nonumber
 I(t) :=& \sum_j |g_j|^2 \int_0^t d\tau_1  \int_0^t d\tau_2  G^*(t-\tau_1) G(t-\tau_2) \times \\
 & \quad\quad  \times e^{i \omega_j(\tau_1-\tau_2)} \braket{\braket{c_j^\dag c_j}}_0\; , \\ \nonumber
 J(t) :=& \sum_j g_j^2 \int_0^t d\tau_1  \int_0^t d\tau_2 G(t-\tau_1) G(t-\tau_2) \times \\ \label{eq:J} 
 &\quad \quad\times e^{-i \omega_j(\tau_1+\tau_2)} \braket{\braket{c_j c_j}}_0\; .
\end{align}

To derive the master equation in time local form, we first write down the exact equations of motion for the system moments, so that we can use them to match a suitable master equation ansatz:
\begin{align} 
 \frac{d}{dt} \langle a \rangle_t =& \frac{\dot{G}(t)}{G(t)} \langle a \rangle_t
 +f(t), \label{eq:FA-d-moment-a-app} \\
 \frac{d}{dt} \langle aa \rangle_t =& 2 \frac{\dot{G}(t)}{G(t)} \langle aa \rangle_t
 +2 f(t)\braket{a}_t +2 \frac{\dot{G}(t)}{G(t)} J(t) - \dot{J}(t) , \label{eq:FA-d-moment-aa-app} \\
 \frac{d}{dt} \langle a^{\dagger}a \rangle_t =&  2 \text{Re} \left(\frac{\dot{G}(t)}{G(t)}\right) \langle a^{\dagger}a \rangle_t 
 +f(t) \braket{a^\dag}_t
  +f^*(t) \braket{a}_t \nonumber
 \\
 &- 2 \text{Re} \left(\frac{\dot{G}(t)}{G(t)}\right) I(t) + \dot{I}(t). \label{eq:FA-d-moment-a*a-app}
\end{align}
where
\begin{align}
f(t) = \dot{F}(t)- \frac{\dot{G}(t)}{G(t)} F(t) 
\end{align}
Based on the fact that the microscopic Hamiltonian is quadratic -- thus that the dynamics preserves the Gaussianity of states -- we make the ansatz that the master equation must be quadratic in the creation and annihilation operators $a^\dagger$, $a$ of the central mode. Namely we already assume that the master equation is in the following time-local, generalized Lindblad form
\begin{equation} \label{eq:FA-ansatz-tcl-meq}
 \frac{d}{dt}\rho_S(t) = -i \left[\tilde{K}_S(t),\rho_S(t)\right] + \tilde{\mathcal{D}}_t  [\rho_S(t)],
\end{equation}
where the effective Hamiltonian is at most quadratic
\begin{equation} \label{eq:FA-ansatz-K_S-fano-anderson}
 \tilde{K}_S(t) = \omega_r(t) a^{\dagger}a + if_r(t)a^\dagger - if_r^*(t)a,
\end{equation}
(there could be in principle also terms in $aa$ and $a^\dag a^\dag$, but they actually drop out) and the dissipator is given in terms of creation and annihilation operators as Lindblad operators, though in non-diagonal form
\begin{align} \label{eq:FA-ansatz-dissipator-fano-anderson}
  \tilde{\mathcal{D}}_t[\rho_S] =&  
 d_1(t) \left[a\rho_Sa^{\dagger}-\frac{1}{2}\left\{a^{\dagger}a,\rho_S\right\}\right] \nonumber \\&+d_2(t) \left[a^{\dagger}\rho_Sa-\frac{1}{2}\left\{aa^{\dagger},\rho_S\right\}\right] \nonumber \\ 
  &  - d_3(t) \left[a\rho_Sa-\frac{1}{2}\left\{aa,\rho_S\right\}\right] \nonumber \\&- d^*_3(t) \left[a^{\dagger}\rho_Sa^{\dagger}-\frac{1}{2}\left\{a^{\dagger}a^{\dagger},\rho_S\right\}\right] \; .
\end{align}
This dissipator can be understood as already satisfying minimal dissipation according to Sec.~\ref{sec:DEQT}. In the ansatz we have given five time dependent coefficients $\omega_r(t)$, $f_r(t)$, $d_1(t)$, $d_2(t)$ and $d_3(t)$; notice that the coefficients $\omega_r(t)$, $d_1(t)$ and $d_2(t)$ must be real as a consequence of Hermiticity preservation of the generator. According to the conjectured master equation, we find the following equations of motion for the moments in terms of the five coefficients:
\begin{align}\label{eq:FA-ansatz-a}
 \frac{d}{dt} \langle a \rangle_t =& \left[ -i\omega_r(t) + \frac{d_2(t)-d_1(t)}{2} \right] \langle a \rangle_t + f_r(t), \;  \\\label{eq:FA-ansatz-aa}
 \frac{d}{dt} \langle aa \rangle_t =&  \left[-2i\omega_r(t) + d_2(t) - d_1(t)\right] \langle a a \rangle_t  \nonumber \\ &+2  f_r(t) \langle a \rangle_t  + d_3(t)\\\label{eq:FA-ansatz-ada}
 \frac{d}{dt} \langle a^{\dagger}a \rangle_t =&   \left[d_2(t)-d_1(t)\right] \langle a^\dagger a \rangle_t + d_2(t)\nonumber \\ &+ f_r(t) \langle a \rangle_t^* + f_r^*(t) \langle a \rangle_t \;. 
\end{align}
Comparing these equations to \eqref{eq:FA-d-moment-a-app}-\eqref{eq:FA-d-moment-a*a-app}, we find that 
\begin{equation}\label{eq:FA-omega-r}
 \omega_r(t) =  -\Im \left(\frac{\dot{G}(t)}{G(t)}\right) \; ,
\end{equation}
while the complex parameters are
\begin{align}
f_r(t) &= f(t) \\
d_3 (t) &= \dot{J}(t) - 2 \frac{\dot{G}(t)}{G(t)} J(t)  
\end{align}
By defining now a generalized transition rate $\gamma(t)$ and a generalized average bath excitation number $N(t)$
\begin{equation} \label{eq:FA-def-gamma}
 \gamma(t) := - 2 \Re \left( \frac{\dot{G}(t)}{G(t)} \right) \; , \quad 
 N(t) := I(t) + \frac{\dot{I}(t)}{\gamma(t)} \; ,
\end{equation}
we get, for the remaining coefficients
\begin{equation} 
 d_1(t) = d_2(t) + \gamma(t) = \gamma(t)(N(t)+1)\; . \label{eq:FA-d1t}
\end{equation}
Combining these results leads to the exact master equation for the central mode in the Fano-Anderson model \eqref{eq:FA-TCL}.

\section{Renormalized driving}\label{app:FA-driving}
In the context of the Fano-Anderson model we can also see, exactly and explicitly, what happens when we impose an external driving protocol on the reduced system while it is connected to the environment. In light of the frequency renormalization of the bare frequency of the system Hamiltonian as a consequence of interaction, a good question to ask is whether an arbitrary coupling to an environment modifies the effect of the external work protocol in terms of work contributions. We will see that this is in general the case in the Fano-Anderson model, where the driving function undergoes renormalization, similarly to the central mode frequency. 

We start with the same assumptions as in the main text, namely factorizing total initial state with a thermal initial environmental state, and modify the global Hamiltonian of the model \eqref{eq:FA-total-Hamiltonian} to allow for external driving on the central mode, i.e. substituting $H_S$ with
\begin{equation}\label{eq:FA-driving-ham}
H_S(t) = \omega_0 a^\dag a +i l(t) a^\dag -i l^*(t) a \; ,
\end{equation}
where $l(t)$ is taken to be a completely arbitrary complex function of time. Then, the solution of the model follows just as in Appendix~\ref{app:FA-master-equation}, with modified Heisenberg equations of motion
\begin{align}
\dot{a}(t) &= -i \omega_0 a(t) +l(t) -i \sum_j g_j c_j(t) \; , \\
\dot{c}_j(t) &= -i \omega_j c_j(t) -i  g_j^* a(t) \; ,
\end{align}
where the equation for the bath modes is identical to the non-driven case, due to the fact that the driving protocol only acts on the system. As before, we find an integro-differential equation for the system mode
\begin{equation}\label{eq:FA-driving-integrodiff-a}
\frac{\mathd}{\mathd t}a(t)  +i \omega_0 a(t) + \int_0^t \mathd \tau \mathcal{K}(t-\tau) a(\tau) = c(t) +l(t) \; .
\end{equation}
with the same memory kernel \eqref{eq:FA-memory-kernel-2-app} as before, but where the driving function $l(t)$ is added to the inhomogeneity \eqref{eq:FA-def-c}. Since the homogeneous part of the differential equation is equivalent to the undriven case, we can use the exact same Green's function $G(t)$ as in Appendix~\ref{app:FA-master-equation}, see eq.~\eqref{eq:FA-green-function-app}. This will make it much easier to compare upcoming results to the undriven case; notice, however, that this is not true for all driving protocols --- it wouldn't be the case, for example, if the time dependency of the system Hamiltonian would have been realized by changing directly the central mode frequency with time. For our chosen driving terms, though, the exact Heisenberg evolution for $a$ is given by
\begin{equation} \label{eq:FA-driving-Heisenberg-solution}
 a(t) = G(t) a(0) + \int_0^t \mathd \tau G(t-\tau) c(\tau) + F(t) \; ,
\end{equation}
where we have defined
\begin{equation}
F(t) = \int_0^t \mathd \tau G(t-\tau) l(\tau) \; .
\end{equation}

We proceed exactly as we did in the undriven case,  deriving from the Heisenberg evolution \eqref{eq:FA-driving-Heisenberg-solution} exact expressions for the first and second moments of the open system mode, which are sufficient to describe the whole dynamics, and which are now given also in terms of the driving function $l(t)$:
\begin{align} 
 \langle a \rangle_t =& G(t) \langle a \rangle_0 +F(t), \label{eq:FA-driving-moment-a} \\
 \langle aa \rangle_t =& G^2(t) \langle aa \rangle_0 +2F(t)G(t)  \langle a \rangle_0 + F^2(t) \, \label{eq:FA-driving-moment-aa} \\
 \langle a^{\dagger}a \rangle_t =& |G(t)|^2 \langle a^{\dagger}a \rangle_0 + I(t) + F(t)G^*(t)  \langle a \rangle_0^* \nonumber \\ &+ F^*(t)G(t)  \langle a \rangle_0 + |F(t)|^2 \label{eq:FA-driving-moment-a*a}
\end{align}
where the noise integral $I(t)$ is still given by \eqref{eq:FA-noise-integral}. Again from the above we derive the exact equations of motion for the relevant system moments, with the goal to compare their terms with those given by a master equation ansatz. They read:
\begin{align} 
 \frac{d}{dt} \langle a \rangle_t =& \frac{\dot{G}(t)}{G(t)} \langle a \rangle_t +\dot{F}(t) - \frac{\dot{G}(t)}{G(t)} F(t) \; , \label{eq:FA-driving-d-moment-a} \\
 \frac{d}{dt} \langle aa \rangle_t =& 2 \frac{\dot{G}(t)}{G(t)} \langle aa \rangle_t + 2\left[ \dot{F}(t) - \frac{\dot{G}(t)}{G(t)} F(t) \right]  \langle a \rangle_t    \;, \label{eq:FA-driving-d-moment-aa} \\ \nonumber
 \frac{d}{dt} \langle a^{\dagger}a \rangle_t =&  2 \text{Re} \left(\frac{\dot{G}(t)}{G(t)}\right) \langle a^{\dagger}a \rangle_t - 2 \text{Re} \left(\frac{\dot{G}(t)}{G(t)}\right) I(t) \\ & + \dot{I}(t)
  + \left[ \dot{F}(t) - \frac{\dot{G}(t)}{G(t)} F(t) \right]  \langle a \rangle_t^* \nonumber \\ & + \left[ \dot{F}^*(t) - \frac{\dot{G}^*(t)}{G^*(t)} F^*(t) \right]  \langle a \rangle_t \; .\label{eq:FA-driving-d-moment-a*a}
\end{align}

Now we make again an ansatz for the master equation. Since the new driven Hamiltonian is still at most quadratic, with a shape that fits our previous ansatz \eqref{eq:FA-ansatz-K_S-fano-anderson}, we keep the exact same ansatz for both parts of the master equation, namely eq. \eqref{eq:FA-ansatz-K_S-fano-anderson} for $K_S$ and eq.  \eqref{eq:FA-ansatz-dissipator-fano-anderson} for the dissipator. The equations for the moments $\langle a \rangle_t$, $\langle aa \rangle_t$ and $\langle a^\dag a \rangle_t$ are then still given by eqns. \eqref{eq:FA-ansatz-a}-\eqref{eq:FA-ansatz-ada}. Comparing their coefficients $\omega_r$, $f_r$, and $d_{1,2,3}$ with the new equations \eqref{eq:FA-driving-d-moment-a}, \eqref{eq:FA-driving-d-moment-a} and \eqref{eq:FA-driving-d-moment-a}, we find a perfect correspondence with the earlier coefficients $\omega_r(t)$, $\gamma(t)$, $d_1(t)$, $d_2(t)$ and $N(t)$ ($d_3(t)=0$ because the initial state of the environment is thermal).
The master equation is therefore identical to the undriven case except for the driving force parameter
\begin{align}\label{eq:FA-driving-ren-force}
f_r(t) =& \dot{F}(t) -\frac{\dot{G}(t)}{G(t)} F(t) 
\nonumber \\ =& l(t) + \int_0^t \mathd \tau  \left[  \frac{\dot{G}(t-\tau)}{G(t-\tau)} - \frac{\dot{G}(t)}{G(t)} \right] G(t-\tau) l(\tau) \; .
\end{align}

This equation shows that there is a time-dependent shift in the original driving force that depends on the shape of $G(t)$. This means that, in general, a driving protocol acting on an open system may be affected by the coupling with the environment, thus one should be careful when considering such scenarios. Here, for example, the driving force gets renormalized and reshaped in the effective Hamiltonian (we expect that similar effects appear in classical systems \cite{Glatzel2022}). We can also observe, by writing the new driving force in terms of other renormalized parameters, namely
\begin{align}\label{eq:FA-driving-ren-force-2}
f_r(t) =& l(t) -i\int_0^t \mathd \tau  [\omega_r(t-\tau) - \omega_r(t)] G(t-\tau) l(\tau) \nonumber \\
&- \int_0^t \mathd \tau \frac{1}{2}[\gamma(t-\tau) - \gamma(t)] G(t-\tau) l(\tau) \; ,
\end{align}
that if both $\omega_r$ and $\gamma$ are time independent, then the force in the Hamiltonian is not renormalized, and the driving protocol remains unaltered. However, even if the renormalized frequency $\omega_r$ is time independent -- which would lead to zero work contribution in the undriven case -- it is still possible that the eventual time-dependency of $\gamma$ affects the driving renormalization non-trivially. The take-home message is: in arbitrarily coupled open systems, even if in the absence of external work protocols the environment does not perform work on the system, it does \emph{not} follow that an additional external driving would be left unmodified. Thus, extra care should always be taken to ensure that the real effective work protocol, with possible modifications emerging from the interaction with the bath, is adequately treated.

\section{Relaxation to equilibrium and to nonequilibrium steady states}\label{app:FA-relaxation}
In this section, we report several details needed to understand the relaxation of the open system to a unique steady state in the long time limit.

\subsection{Final excitation number}\label{app:FA-relaxation-eq}
To evaluate in general the final steady state for a thermal case, we take the limit to infinity of the noise integral \eqref{eq:FA-noise-integral}. We can also perform a change of variables and exploit the fact that the correlation function $ \langle c^{\dagger}(\tau_1) c(\tau_2) \rangle $ depends only on the time difference $(\tau_1-\tau_2)$ to find
\begin{equation}\label{eq:FA-noise-infty-1}
I(\infty) = \int_0^\infty d t_1  \int_0^\infty d t_2 G^*(t_1) G(t_2) \langle c^{\dagger}(t_2) c(t_1) \rangle \; .
\end{equation}
The correlation function reads
\begin{align}
\langle c^{\dagger}(t_2) c(t_1) \rangle =& \sum_{i,j} g_i^* g_j e^{i\omega_i t_2} e^{-i\omega_jt_1} \langle c_i^{\dagger}(0) c_j(0) \rangle \nonumber \\=&  \sum_{j} |g_j |^2e^{i\omega_j (t_2-t_1)} \frac{1}{e^{\beta \omega_j}-1} \; ,
\end{align}
where we have used $\langle c_i^{\dagger}(0) c_j(0) \rangle = \delta_{ij}n_j(0)$ with $n_j(0)$ the Planck distribution for energy $\omega_j$ and inverse temperature $\beta$. Taking the continuum limit we find
\begin{equation}
\langle c^{\dagger}(t_2) c(t_1) \rangle = \int_0^\infty\mathd \omega J(\omega) e^{i\omega (t_2-t_1)} \frac{1}{e^{\beta \omega}-1}  \; ,
\end{equation}
which, inserted into \eqref{eq:FA-noise-infty-1} gives the final steady state excitation number in terms of the spectral density and the Laplace transform of $G(t)$, namely
\begin{equation}\label{eq:FA-noise-infty-2}
\langle a^\dag a \rangle_\infty = \int_0^\infty \mathd \omega J(\omega) \frac{1}{e^{\beta \omega}-1}  |\hat{G}(-i\omega)|^2  \; .
\end{equation}
The Laplace transform $\hat{G}$ is given by eq. \eqref{eq:FA-green-function} in terms of the Laplace transform of the memory kernel \eqref{eq:FA-memory-kernel-2}:
\begin{equation}
\hat{G}(-i\omega) = \frac{1}{i(\omega_0-\omega)+\hat{\mathcal{K}}(-i\omega)} \; , 
\end{equation}
with
\begin{equation} 
\hat{\mathcal{K}}(z) = \int_0^\infty \mathd \omega' \frac{J(\omega')}{z+i\omega'} \; .
\end{equation}
In turn, this is given in terms of a principal value integral
\begin{equation}\label{eq:FA-Laplace-K-iw}
\hat{\mathcal{K}}(-i\omega)= \pi J(\omega) + i \Delta(\omega) \; , 
\end{equation}
\begin{equation}
\Delta(\omega) = \mathcal{P} \int_0^\infty \mathd \omega' \frac{J(\omega')}{\omega - \omega'} \; ,
\end{equation}
giving finally
\begin{equation}\label{eq:FA-Laplace-G}
\hat{G}(-i\omega) = \frac{1}{i(\omega_0+\Delta(\omega)-\omega)+\pi J(\omega)} \; ,
\end{equation}
and the steady state excitation number
\begin{align}\label{eq:FA-noise-infty-3}
\langle a^\dag a \rangle_\infty = \int_0^\infty \mathd \omega J(\omega)& \frac{1}{e^{\beta \omega}-1} \times \nonumber \\
&\times \frac{1}{[\omega_0+\Delta(\omega)-\omega]^2+\pi^2 J^2(\omega)}  \; .
\end{align}
We will use this result in the next section to prove that the above is equivalent to the equilibrium expectation value --- namely considering system and environment in a global Gibbs state at the initial bath temperature.

\subsection{Proof of thermal equilibrium at bath temperature}
We have already proven that the model relaxes to a unique final steady state; we will now show that this state is equivalent to the mean-force state obtained by tracing out a global thermal state:
\begin{equation}
\rho_S^*= \Tr_E \left\{ \frac{e^{-\beta H_{SE}}}{Z_{SE}}\right\}  \;,
\end{equation}
with $\beta$ still the initial inverse temperature of the environment. To prove this it is sufficient to show that the average values of all relevant moments at long times, namely eqs. \eqref{eq:FA-moment-a-long}-\eqref{eq:FA-moment-a*a-long}, coincide with the average taken over a global Gibbs state. The expectation value of $a$ and $aa$ are zero for a Gibbs state of the total Hamiltonian, so the only value left to check is that of 
\begin{equation}
\langle a^\dag a \rangle_{\text{eq}} := \Tr \left\{ a^\dag a \otimes \mathbb{I} \frac{e^{-\beta H_{SE}}}{Z_{SE}}\right\} \; .
\end{equation}
For this purpose we make use of the equilibrium fluctuation dissipation theorem, for which the expectation value of the position operator $X = (a^\dag+a)/\sqrt{2}$ squared is given in terms of the Fourier transform of the response function 
\begin{equation}\label{eq:FA-response-function}
\chi(t) := i \theta(t)\langle [X(t), X(0)] \rangle = \frac{i}{2} \theta(t) \left( G(t)- G^*(t) \right) \; ,
\end{equation}
namely
\begin{equation}\label{eq:FA-x^2}
\langle X^2 \rangle_{\text{eq}} =  \frac{1}{2\pi} \int_{-\infty}^{+\infty} \mathd \omega \coth\left(\frac{\beta\omega}{2}\right)\Im \{\tilde{\chi}(\omega)\} \; , 
\end{equation}
where $\tilde{\chi}(\omega)$ denotes the Fourier transform of $\chi(t)$. Evaluating the above is useful because of the easily proven relation
\begin{equation}\label{eq:FA-neq-1}
\langle a^\dag a \rangle_{\text{eq}} = \langle X^2 \rangle_{\text{eq}} - \frac{1}{2} =\frac{1}{\pi}  \int_{0}^{\infty} \mathd \omega (2 n(\omega)+1)\Im\{\tilde{\chi}(\omega)\} \; ,
\end{equation}
where we have used the fact that the integrand in \eqref{eq:FA-x^2} is symmetric in $\omega$ and have inserted $\coth({\beta\omega/2}) = 2 n(\omega)+1$.
Using \eqref{eq:FA-response-function}, the Fourier transform of ${\chi}$ is given by 
\begin{equation}
\tilde{\chi}(\omega) =  \int_{-\infty}^{+\infty} \mathd t e^{i \omega t} \chi(t) =  \frac{i}{2} \left[ \hat{G}(-i \omega) - \hat{G}^*(i \omega) \right] \; ,
\end{equation}
such that its imaginary part is given by 
\begin{equation}
\Im\{\tilde{\chi}(\omega)\} =  \frac{1}{2} \left[ \Re\{\hat{G}(-i \omega)\} - \Re\{\hat{G}(i \omega)\}\right] \; .
\end{equation} Here, the second term vanishes and, recalling \eqref{eq:FA-Laplace-G}, we are left with
\begin{equation}
\Im\{\tilde{\chi}(\omega)\} = \frac{1}{2}\frac{\pi J(\omega)}{[\omega_0+\Delta(\omega)-\omega]^2+\pi^2 J^2(\omega)}\; .
\end{equation}
Notice now that the following integral, whenever all poles of $\hat{G}(z)$ have negative real part, is equal to one:
\begin{equation}
 \int_{0}^{\infty} \mathd \omega \frac{ J(\omega)}{[\omega_0+\Delta(\omega)-\omega]^2+\pi^2 J^2(\omega)} =1 \; ;
\end{equation}
using all the above in \eqref{eq:FA-neq-1} we finally arrive at
\begin{equation}\label{eq:FA-neq-equiv}
\langle a^\dag a \rangle_{\text{eq}} =\int_{0}^{\infty} \mathd \omega J(\omega)n(\omega) \frac{1}{[\omega_0+\Delta(\omega)-\omega]^2+\pi^2 J^2(\omega)}  \; ,
\end{equation}
which is identical to the steady state value \eqref{eq:FA-noise-infty-3}. This shows that whenever $G(t)$ has negative Laplace transform poles (which in turn implies that $G(t)$ vanishes in the long time limit, as already assumed) then the final steady state is equivalent to the state which shows the system's perspective of global thermal equilibrium.

\subsection{Approach to a nonequilibrium steady state}\label{app:FA-NESS}

From the exact evolution of the moments \eqref{eq:FA-moment-a}-\eqref{eq:FA-moment-a*a}, the limit of the Green function $G(t)$ decaying to zero at long times leads to the disappearance of all dependence on the initial system state just like in the thermal case, and the noise integral $I(t)$ converges to the same value $I(\infty)$ of \eqref{eq:FA-noise-infty-3-main}. 

To understand what happens to the displacement term $F(t)$ in the long time limit, we rewrite it as
\begin{align}
F(t) =& -i \sum_j g_j \alpha_j e^{-i\omega_j t} \int_0^t d\tau G(\tau)e^{i\omega_j \tau} \;.
\end{align}
Then, in the limit where $G$ has decayed we can take the limit to infinity in the integral to obtain
\begin{align}
\overline{F}(t) =& -i \sum_j g_j \alpha_j e^{-i\omega_j t} \int_0^\infty d\tau G(\tau)e^{i\omega_j \tau} = \nonumber \\
=&-i \sum_j g_j \hat{G}(-i\omega_j) \alpha_j e^{-i\omega_j t}\; ,
\end{align}
which gives the final expression for the nonequilibrium steady state displacement. 

Since the long time limit of $\dot{G}(t)/G(t)$ is given by the steady values of $\gamma$ and $\omega_r$ in the undriven case,
\begin{equation}
\lim_{t\rightarrow \infty} \frac{\dot{G}(t)}{G(t)} = -\left(\frac{\overline{\gamma}}{2} + i \overline{\omega}_r\right)
\end{equation}
one obtains the long time limit driving force \eqref{eq:FA-NESS-force} simply from
\begin{equation}
\overline{f}(t) = \dot{\overline{F}}(t) + \left(\frac{\overline{\gamma}}{2} + i \overline{\omega}_r\right)\overline{F}(t) \; .
\end{equation}

To check that the unitary evolution of the state $\overline{\rho}_S(t)$ is also described by the long-time limit master equation \eqref{eq:FA-NESS-long-time-me}, we first rewrite equation \eqref{eq:FA-NESS-evolution} as a commutator with a Hamiltonian:
\begin{equation}
\frac{d}{dt}\overline{\rho}_S(t) = -i [\overline{H}(t), \overline{\rho}_S(t)] \; ,
\end{equation}
where the Hamiltonian generating the unitary evolution of the NESS is given by $\overline{H}(t)= i \dot{D}_t D^{\dag}_t$, thus in turn by 
\begin{equation}
\overline{H}(t)= i \dot{\overline{F}}(t) a^{\dag} - i \dot{\overline{F}}^*(t) a \; . 
\end{equation}
It should hold that $\overline{\mathcal{L}}_t[\overline{\rho}_S(t)] =  -i [\overline{H}(t), \overline{\rho}_S(t)]$. One can see that, due to the properties of displacement operators on $a$ and $a^\dag$ and that $\overline{\mathcal{D}}[\rho^{G_r}_S]= 0$, the dissipator part acting on the NESS gives itself a Hamiltonian contribution
\begin{equation}
\overline{\mathcal{D}}[\overline{\rho}_S(t)] = -i [\overline{H}_D(t), \overline{\rho}_S(t)] \; ,
\end{equation}
with 
\begin{equation}
\overline{H}_D(t)= i \frac{\overline{\gamma}}{2}\left(\overline{F}^*(t) a - \overline{F}(t) a^{\dag}\right) \; ,
\end{equation}
which leads to $\overline{\mathcal{L}}_t[\overline{\rho}_S(t)] =-i[\overline{K}_S(t)+\overline{H}_D(t),\overline{\rho}_S(t)]$. It is easy to check that 
\begin{equation}
\overline{K}_S(t)+\overline{H}_D(t) = \overline{\omega}_r (a^\dag - \overline{F}^*(t))(a - \overline{F}(t)) + \overline{H}(t) \; ,
\end{equation}
and that 
\begin{equation}
[\overline{\omega}_r (a^\dag - \overline{F}^*(t))(a - \overline{F}(t)) ,  \overline{\rho}_S(t)] = D_t[\overline{\omega}_r a^\dag a ,  \rho^{G_r}_S]D^{\dag}_t =0\; ,
\end{equation}
which proves the equivalence between the two evolutions on this state.

\section{Born-Markov approximation}\label{app:FA-Born-Markov}
To obtain the Born-Markov limit of the model in the thermal case, we assume weak system-bath coupling by isolating a coupling parameter $\lambda$ from the interaction Hamiltonian.
This leads to a spectral density of second order in the coupling, namely $\lambda^2 J(\omega)$. Then, the still exact integro-differential equation \eqref{eq:FA-green-function} for the Green function $G(t)$, where we now explicitly write out the coupling parameter $\lambda$, gives
\begin{equation}
 \frac{d}{dt}G(t) + i\omega_0 G(t) + \lambda^2  \int_0^t \mathd \tau \mathcal{K}(t-\tau) G(\tau) = 0.
\end{equation}
From the above we can evaluate $G(t)$ at different orders of the coupling to obtain a second-order approximation. Because of the structure of the differential equation, we make the ansatz that only terms of even order in $\lambda$ appear in $G(t)$, so that we can write it as
\begin{equation}
G(t) = G_0(t) +\lambda^2 G_2(t) +\lambda^4 G_4(t) + ... \; .
\end{equation}
Proceding order by order in solving the differential equation gives the zeroth order equation for $G_0$ and its solution:
\begin{equation}
 \frac{d}{dt}G_0(t) + i\omega_0 G_0(t) = 0 \quad \implies \quad G_0(t) = e^{-i \omega_0 t}\; ,
\end{equation}
with $G_0(0)=1$, and consequently the second order ones for $G_2$:
\begin{equation}
 \frac{d}{dt}G_2(t) + i\omega_0 G_2(t) +  \int_0^t \mathd \tau \mathcal{K}(t-\tau) G_0(\tau) = 0 \; ,
\end{equation}
with $G_2(0)=0$. This implies
\begin{equation}
G_2(t) = - e^{-i \omega_0 t} \int_0^t \mathd \tau   \int_0^\tau \mathd s \mathcal{K}(\tau-s) e^{-i \omega_0 (\tau - s)} \; .
\end{equation}

We want to use these results to evaluate the full second order Green function and its differential equation, in order to obtain an approximation of the relevant parameters in the master equation. They are:
\begin{align}
G^{(2)}(t) = e^{-i \omega_0 t} \left[ 1- \lambda^2 \int_0^t \mathd \tau   \int_0^\tau \mathd s \mathcal{K}(\tau-s) e^{-i \omega_0 (\tau - s)} \right]\; ,
\end{align}
\begin{align}
\dot{G}^{(2)}(t) =&  - i\omega_0 G^{(2)}(t)  \\ \nonumber & - \lambda^2 e^{-i \omega_0 t}  \int_0^t \mathd \tau \mathcal{K}(t-\tau) e^{-i \omega_0 (t-\tau)}   \; .
\end{align}
Inserting the expression of the memory kernel in terms of the spectral density gives the following equation for the central second order quantity
\begin{equation}\label{eq:FA-GdotG-second-order}
\frac{\dot{G}^{(2)}(t)}{G^{(2)}(t)} = -i \omega_0 -\lambda^2 \int_0^\infty \mathd \omega J(\omega) \int_0^t \mathd \tau    e^{-i (\omega-\omega_0) \tau} \; , 
\end{equation}
which determines the second order expansion of the parameters $\gamma(t)$ and $\omega_r(t)$. 

The Markovian approximation is employed by taking the limit $t\rightarrow \infty$ in the integral to obtain
\begin{equation}
\frac{\dot{G}^{(2)}(t)}{G^{(2)}(t)} = -i \omega_0 - i\lambda^2 \mathcal{P} \int_0^\infty \mathd \omega \frac{J(\omega)} {(\omega_0-\omega)} - \lambda^2\pi J(\omega_0) \; ,
\end{equation}
where $\mathcal{P}$ denotes the Cauchy principal value. The real and imaginary part give the Markovian coefficients in the master equation at second order:
\begin{align}
\gamma_0 =& 2 \lambda^2 \pi J(\omega_0)  \; ,  \\
\omega_r =& \omega_0 + \lambda^2 \delta \omega =  \omega_0 + \lambda^2 \mathcal{P} \int_0^\infty \mathd \omega \frac{J(\omega)} {(\omega_0-\omega)}    \; ,
\end{align}
which are time independent as a consequence of the Markov approximation. 

Also in the case of a linear driving force as explored in Appendix~\ref{app:FA-driving}, we find that the driving force does not get renormalized. Indeed, recalling the results of Appendix~\ref{app:FA-driving} for the renormalization of the driving force, namely eq. \eqref{eq:FA-driving-ren-force-2}, we see that finding time-independent rate $\gamma$ and frequency $\omega_r$ in the undriven model implies that the driving force is left unaltered. Thus, in the case where the Born-Markov approximation can be applied, there is no modification of a driving protocol of the form \eqref{eq:FA-driving-ham}, such that the work done on the system is fully determined by the external driving force, while the environment is only providing heat exchange. However, we remark again that this result might not hold for all types of driving -- for example, it might fail for quadratic driving as opposed to the proposed linear driving.

\bibliography{biblio}

\end{document}